\documentclass[structabstract]{aa}

\usepackage{url}
\usepackage[breaklinks=true]{hyperref}
\usepackage{twoopt}
\usepackage[english]{babel}          

\usepackage{natbib}
\bibpunct{(}{)}{;}{a}{}{,} 

\usepackage[utf8]{inputenc}
\usepackage{siunitx}
\usepackage{amsmath, amssymb, amsfonts}
\usepackage[farskip=0pt,caption=false]{subfig}

\usepackage{multirow,bigstrut,ctable}
\usepackage[normalem]{ulem}
\usepackage{rotating}
\usepackage[varg]{txfonts} 
\usepackage{threeparttable}

\def \deg         {\text{$^{\circ}$}}

\def \arcsec      {\text{$^{\prime\prime}$}}

\def \mjybeam     {mJy\,beam$^{-1}$}
\def \mujybeam    {$\mathrm{\mu}$Jy\,beam$^{-1}$}

\newcommand{\Hii}{\text{H\textsc{ii}}}
\newcommand{\Hi}{\text{H\textsc{i}}}

\newcommand{\beam}[2]{{#1}\arcsec$\times${#2}\arcsec}

\def \lol {LoLSS}

\begin{document}

\title{The LOFAR LBA Sky Survey}
\subtitle{I. Survey description and preliminary data release}
\titlerunning{LOFAR LBA sky survey I}

\author{F.~de~Gasperin\inst{1,2}
\and W.~L.~Williams\inst{3}
\and P.~Best\inst{4}
\and M.~Br\"uggen\inst{1}
\and G.~Brunetti\inst{2}
\and V.~Cuciti\inst{1}
\and T.~J.~Dijkema\inst{5}
\and M.~J.~Hardcastle\inst{6}
\and M.~J.~Norden\inst{5}
\and A.~Offringa\inst{5}
\and T.~Shimwell\inst{3,5}
\and R.~van~Weeren\inst{3}
\and D.~Bomans\inst{7}
\and A.~Bonafede\inst{2,8}
\and A.~Botteon\inst{3}
\and J.~R.~Callingham\inst{3,5}
\and R.~Cassano\inst{2}
\and K.~T.~Chy\.zy\inst{9}
\and K.~L.~Emig\inst{3,10}\thanks{K. L. Emig is a Jansky Fellow of the National Radio Astronomy Observatory}
\and H.~Edler\inst{1}
\and M.~Haverkorn\inst{11}
\and G.~Heald\inst{12}
\and V.~Heesen\inst{1}
\and M.~Iacobelli\inst{5}
\and H.~T.~Intema\inst{3}
\and M.~Kadler\inst{13}
\and K.~Ma\l{}ek\inst{14}
\and M.~Mevius\inst{5}
\and G.~Miley\inst{3}
\and B.~Mingo\inst{15}
\and L.~K.~Morabito\inst{16,17}
\and J.~Sabater\inst{4}
\and R.~Morganti\inst{5,18}
\and E.~Orr\'u\inst{5}
\and R.~Pizzo\inst{5}
\and I.~Prandoni\inst{2}
\and A.~Shulevski\inst{3,19}
\and C.~Tasse\inst{20,21}
\and M.~Vaccari\inst{2,22}
\and P.~Zarka\inst{23}
\and H.~R\"ottgering\inst{3}
}

\authorrunning{F.~de~Gasperin et al.}

\institute{\tiny
Hamburger Sternwarte, Universit\"at Hamburg, Gojenbergsweg 112, D-21029, Hamburg, Germany
\and INAF - Istituto di Radioastronomia, via P. Gobetti 101, 40129, Bologna, Italy
\and Leiden Observatory, Leiden University, P.O.Box 9513, NL-2300 RA, Leiden, The Netherlands
\and Institute for Astronomy, University of Edinburgh, Royal Observatory, Blackford Hill, Edinburgh, EH9 3HJ, UK
\and ASTRON, the Netherlands Institute for Radio Astronomy, Postbus 2, 7990 AA, Dwingeloo, The Netherlands
\and Centre for Astrophysics Research, University of Hertfordshire, College Lane, Hatfield AL10 9AB, UK
\and Ruhr-Universität Bochum, Universitätsstr 150/NA7, 44801 Bochum, Germany
\and DIFA - Universit\'a di Bologna, via Gobetti 93/2, I-40129 Bologna, Italy
\and Astronomical Observatory, Jagiellonian University, ul. Orla 171, 30-244, Krak\'ow, Poland
\and National Radio Astronomy Observatory, 520 Edgemont Road, Charlottesville, VA 22903-2475, USA 
\and Department of Astrophysics/IMAPP, Radboud University, PO Box 9010, NL-6500 GL Nijmegen, the Netherlands
\and CSIRO Astronomy and Space Science, PO Box 1130, Bentley WA 6102, Australia
\and Institut f\"ur Theoretische Physik und Astrophysik, Universit\"at W\"urzburg, Emil-Fischer-Str. 31, 97074 W\"urzburg, Germany
\and National Centre for Nuclear Research, ul. Pasteura 7, 02-093, Warsaw, Poland
\and School of Physical Sciences, The Open University, Walton Hall, Milton Keynes MK7 6AA, UK
\and Centre for Extragalactic Astronomy, Department of Physics, Durham University, Durham DH1 3LE, UK 
\and Institute for Computational Cosmology, Department of Physics, University of Durham, South Road, Durham DH1 3LE, UK
\and Kapteyn Astronomical Institute, University of Groningen, P.O. Box 800, 9700 AV Groningen, The Netherlands
\and Anton Pannekoek Institute for Astronomy, University of Amsterdam, Postbus 94249, 1090 GE Amsterdam, The Netherlands
\and GEPI\&USN, Observatoire de Paris, CNRS, Universit\'e Paris Diderot, 5 place Jules Janssen, 92190 Meudon, France
\and Centre for Radio Astronomy Techniques and Technologies, Rhodes University, Grahamstown 6140, South Africa
\and Dep. of Physics \& Astronomy, University of the Western Cape, Robert Sobukwe Road, 7535 Bellville, Cape Town, South Africa
\and LESIA, UMR CNRS 8109, Observatoire de Paris, 92195 MEUDON, France
}

\abstract
{The LOw Frequency ARray (LOFAR) is the only radio telescope that is presently capable of high-sensitivity, high-resolution (i.e. $<1$~\mjybeam{} and $<15$\arcsec) observations at ultra-low frequencies ($<100$ MHz). To utilise these capabilities, the LOFAR Surveys Key Science Project is undertaking a large survey to cover the entire northern sky with Low Band Antenna (LBA) observations.}
{The LOFAR LBA Sky Survey (\lol)  aims to cover the entire northern sky with 3170 pointings in the frequency range  between $42-66$ MHz, at a resolution of 15\arcsec{} and at a sensitivity of 1~\mjybeam{} ($1\sigma$). In this work,  we outline the survey strategy, the observational status, and the calibration techniques. We also briefly describe several of our scientific motivations and present the preliminary public data release.}
{The preliminary images were produced using a fully automated pipeline aimed at correcting all direction-independent effects in the data. Whilst the direction-dependent effects, such as those from the ionosphere, have not yet been corrected, the images presented in this work are still ten times more sensitive than previous available surveys  at these low frequencies.}
{The preliminary data release covers 740 deg$^2$ around the HETDEX spring field region at an angular resolution of 47\arcsec{} with a median noise level of 5~\mjybeam. The images and the catalogue of 25,247 sources have been publicly released. We demonstrate that the system is capable of reaching a root mean square (rms) noise of 1~\mjybeam{} and an angular resolution of 15\arcsec{} once direction-dependent effects are accounted for.}
{\lol{} will provide the ultra-low-frequency information for hundreds of thousands of radio sources, providing critical spectral information and producing a unique data set that can be used for a wide range of science topics, such as the search for high redshift galaxies and quasars, the study of the magnetosphere of exoplanets, and the detection of the oldest populations of cosmic-rays in galaxies, clusters of galaxies, as well as those produced by active galactic nuclei (AGN).}

\maketitle

\section{Introduction}
\label{sec:introduction}

The LOw Frequency ARray \citep[LOFAR;][]{VanHaarlem2013} is a radio interferometric array that operates at very low frequencies ($10-240$~MHz), built with the ambition of performing groundbreaking
imaging surveys \citep{Rottgering2011}. Compared to existing radio telescopes, LOFAR offers the possibility of performing transformational high-resolution surveys thanks to the increase in survey speed resulting from its large field of view (FoV), vast collecting area, and multi-beam capabilities. Two wide-area imaging surveys were designed within its framework:
1) LoTSS \citep[LOFAR Two-metre Sky Survey;][]{Shimwell2017} is a wide area survey at 120--168 MHz that uses the high band antenna (HBA) system of LOFAR; and 2) \lol{} (LOFAR LBA Sky Survey) is the sibling survey of LoTSS carried out in the frequency range 42--66 MHz using the LOFAR Low Band Antenna (LBA) system.

LoTSS and \lol{} are two wide-area surveys led by the LOFAR Survey Key Science Project (SKSP; PI: R\"ottgering). Both surveys aim to cover the northern hemisphere. LoTSS has published its first data release, comprising 424 deg$^2$ of sky and detecting over 320,000 sources \citep{Shimwell2019}. In a select number of regions, where high-quality multi-wavelength data sets are available, the SKSP is also taking longer exposures to achieve significantly higher sensitivities (deep fields). A few of them, observed with the HBA system and reaching a noise level as low as 20~\mujybeam{}, were recently released as part of the LoTSS-deep first data release: Bo\"otes, Lockman, and ELAIS-N1 \citep{Tasse2020, Sabater2020}. In this paper, we focus on the ongoing LOFAR LBA Sky Survey.

LOFAR LBA is currently the only instrument capable of deep (\mjybeam), high-resolution (15\arcsec) imaging at frequencies below 100 MHz. Even into the SKA era, this capability will remain unique to LOFAR. \lol{} is a long-term project and currently,  around 500 deg$^2$ have been observed at the target integration time per pointing of 8 hrs, whilst data from a further 6700 deg$^2$ are being collected with an initial integration time of 3 hrs per pointing.

\begin{figure}
\centering
 \includegraphics[width=0.5\textwidth]{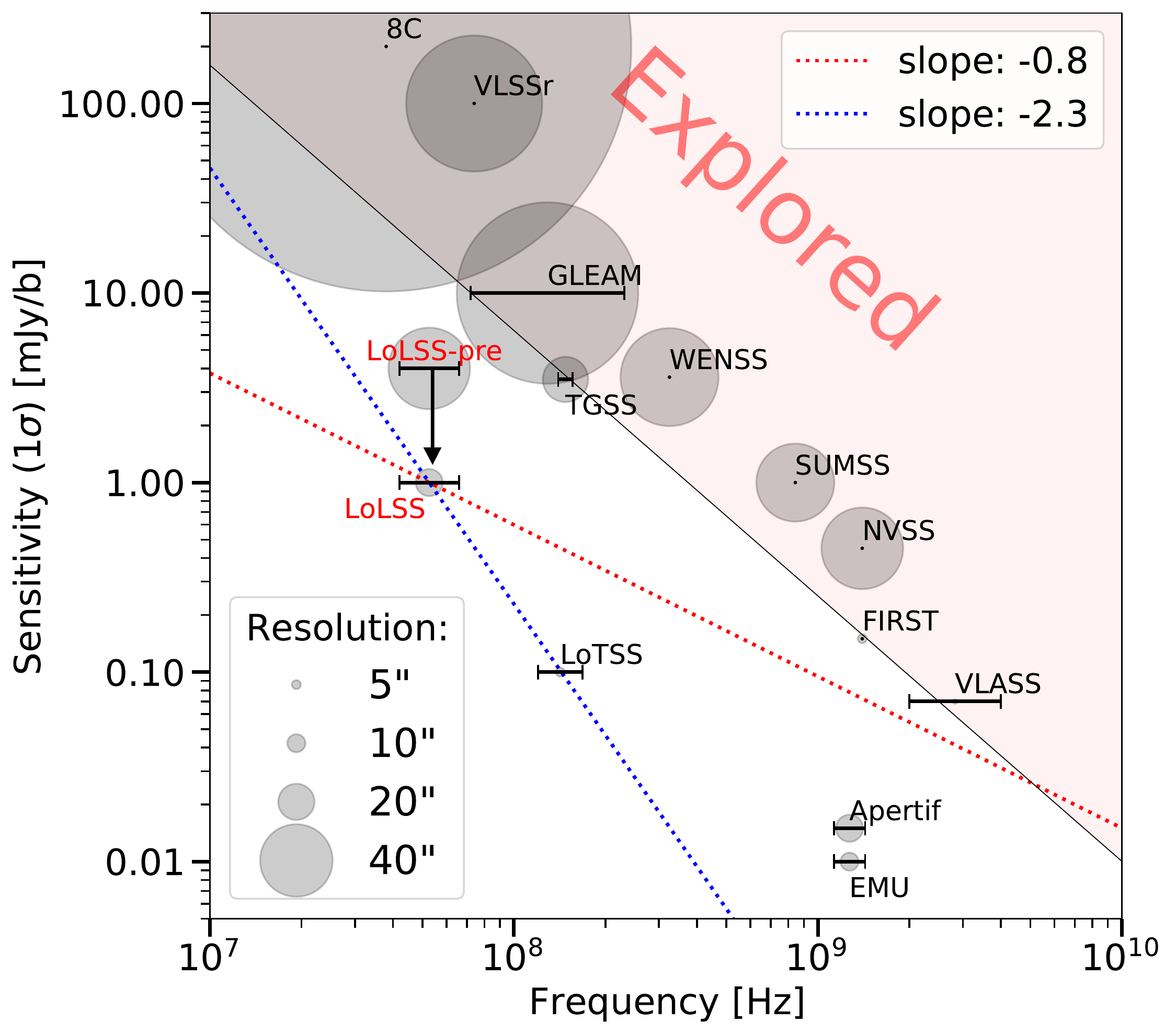}
 \caption{Comparison of sensitivity for a number of completed and ongoing wide-area radio surveys. The diameters of the grey circles are proportional to the survey resolution as shown in the bottom left corner. Data presented in this paper are labelled as `LoLSS-pre', whilst the final \lol{} survey is labelled as `LoLSS'. For sources with a very steep spectral index ($\alpha \lesssim -2.3$), \lol{} will be the most sensitive survey on the market. References: 8C \citep{Rees1990}; GLEAM \citep[GaLactic and Extragalactic All-sky Murchison Widefield Array survey;][]{Hurley-Walker2017}; TGSS ADR1 \citep[TIFR GMRT Sky Survey - Alternative Data Release 1;][]{Intema2017}; VLSSr \citep[VLA Low-frequency Sky Survey redux;][]{Lane2014}; FIRST \citep[Faint Images of the Radio Sky at Twenty Centimetres;][]{Becker1995}; NVSS \citep[1.4 GHz NRAO VLA Sky Survey;][]{Condon1998}; WENSS \citep[The Westerbork Northern Sky Survey;][]{Rengelink1997}; SUMSS \citep[Sydney University Molonglo Sky Survey;][]{Bock1999a}; Apertif (Adams et al. in prep.); EMU \citep[Evolutionary Map of the Universe][]{Norris2011}; VLASS \citep[VLA Sky Survey;][]{Lacy2020}; LoTSS \citep[LOFAR Two-metre Sky Survey;][]{Shimwell2017}.}
 \label{fig:survey}
\end{figure}

\lol{} will open a hitherto unexplored spectral window (Fig.~\ref{fig:survey}), addressing one of the original motivations for the construction of LOFAR. Compared to other ultra-low frequency surveys \citep[VLSSr and GLEAM;][]{Lane2014, Hurley-Walker2017}, \lol{} will be 10-100 times more sensitive and will have an   angular resolution that is 5-10 times higher. For sources with a typical spectral index $\alpha \sim -0.8$ (with $S_{\nu} \propto \nu^{\alpha}$), \lol{} will be more sensitive than the majority of current and planned surveys. For sources with ultra-steep spectra ($\alpha < -2.3$) or sharp spectral cutoffs at low-frequencies, \lol{} will stand as the deepest survey available. In the northern hemisphere, where LoTSS and \lol{} will both cover $2\pi$ steradians, the combination of the two surveys will provide unique insights into the low-frequency spectral index values of a million radio sources.

\section{Science cases}
\label{sec:science_cases}

\lol{} will investigate low-energy synchrotron radiation with a unique combination of high angular resolution and sensitivity, enabling the study of phenomena such as low-efficiency acceleration mechanisms and the detection of old cosmic-ray populations. Studying `fossil' steep-spectrum sources is a requirement for understanding the nature, evolution, and life cycles of synchrotron radio sources. \lol{} will also probe processes that modify the power-law synchrotron spectra at these extreme frequencies, thereby providing new information about certain processes, such as the absorption by ionised gas and synchrotron self-absorption. \lol{} will thus be a unique diagnostic tool for studying both the local and the diffuse medium in a variety of astronomical environments. \lol{} is designed to maximise the synergy with its sibling survey LoTSS. The combination of \lol{} (LBA) and LoTSS (HBA) will produce a unique body of data for the investigation of radio sources at low frequencies, where several new physical diagnostics are available.

\subsection{Distant galaxies and quasars}

Owing to their large luminosities and bright associated emission lines, active galaxies are among the most distant objects observable in the Universe. One of the most efficient techniques for finding high-redshift radio galaxies (HzRGs) and proto-clusters is to focus on ultra-steep spectrum (USS) radio sources \citep{Miley2008, Saxena2018}. One of the ultimate goals of the LOFAR Surveys KSP is to detect $>100$ radio galaxies at $z > 6;$ to enable robust studies of the formation and evolution of high-redshift massive galaxies, black holes, and proto-clusters; and to provide a sufficient number of radio sources within the Epoch of Reionisation to facilitate \Hi{} absorption studies. Combining \lol{} and LoTSS in a large region of the sky will identify USS HzRGs candidates, as well as a set of highly redshifted GHz-peaked sources \citep[peaking at $\sim 100$~MHz;][]{Falcke2004a}, of which $>30$ are expected to be at $z>6$ \citep{Saxena2017}. Distance constraints of the candidates will be enabled by the WEAVE-LOFAR optical spectroscopic survey\footnote{\url{https://ingconfluence.ing.iac.es:8444/confluence//display/WEAV/WEAVE-LOFAR}} \citep{Smith2016}, prior to optical, infrared, and millimetre-wave follow up.

\subsection{Galaxy clusters and large-scale structure}

Being dynamically complex and very large magnetised regions, galaxy clusters are important laboratories for studying the contribution of particle acceleration and transport to cluster evolution \citep[e.g.][]{Brunetti2014}. To date, approximately 100 clusters are known to contain Mpc-sized, steep-spectrum ($\alpha < -1$) synchrotron radio sources that are not associated with individual galaxies. These are classified either as radio haloes, mini-haloes, or radio relics, depending on their location, morphology, and polarisation properties \citep{vanWeeren2019}. \lol{} will detect hundreds of diffuse cluster radio sources out to $z = 1$ \citep{Cassano2010b}. A fundamental prediction of radio halo theories is that many of them should have ultra-steep spectra \citep[$\alpha<-1.5$;][]{Brunetti2008}. The combination of LBA and HBA data will immediately provide resolved spectral index measurements for these sources \citep[see e.g.][ for Abell 2256 and the Toothbrush]{VanWeeren2012f, deGasperin2020a}, whilst high-frequency surveys do not have the required combination of depth, resolution, or coverage to be viable counterparts to LoTSS. For radio relics, both cosmic-ray acceleration at the shock front and their energy loss processes in the post-shock region are poorly understood and tightly linked to the observations at ultra-low frequencies \citep{deGasperin2020a}. \lol{} will thus enable the investigation of the microphysics of cosmic-ray acceleration processes in both radio haloes (turbulence acceleration) and radio relics (shock-induced acceleration). Studies of these processes are expected to place firm constraints on the theoretical models \citep{Brunetti2014}.

Furthermore, recent LOFAR observations have discovered diffuse synchrotron emission from bridges connecting clusters that are still in a pre-merger phase \citep{Botteon2018, Govoni2019, Botteon2020b}. Only few clusters are known to be in such configuration, but data from the eRosita all sky survey (eRASS) will likely increase their number. These observations demonstrate that relativistic electrons and magnetic fields can be generated on very large, cosmological scales that had never been probed before. These pairs of massive clusters sit in largely overdense regions which result from the collapse of cosmic filaments. The resulting bridges are regions where turbulence may amplify magnetic fields and accelerate particles, leading to observable radio emission extending on $3-5$ Mpc scales and with a predicted steep spectral shape \citep[$\alpha \sim -1.5$]{Brunetti2020}. Recent ASKAP early-science observations of the intercluster region of the cluster pair A3391-A3395 have shown that these studies are not easy to carry out at conventional frequencies \citep{Bruggen2020}. Thanks to the ultra-low observing frequency and the high sensitivity to large-scale emission, \lol{} will have the potential to detect emission from such large-scale structures and measure their spectra.

Observations at low frequencies have the ability to trace plasma generated by  activity from active galactic nuclei (AGN) that has been mildly re-energised through compression or other phenomena. Sources of this type can have spectral indices as steep as $\alpha = -4$ (e.g. Gently Re-Energised Tails \citealt{deGasperin2017} or radio Phoenixes \citealt{Mandal2020}). Since the LBA system is nearly ten times more sensitive than HBA for such steep spectra, LOFAR LBA is the only instrument able to efficiently detect this new population of elusive sources. These detections will enable the study of the interaction of radio galaxies and tailed sources with the intra-cluster medium \citep{Bliton1998} as well as the new micro-physics involved in the inefficient re-acceleration of cosmic-rays in diluted plasmas \citep{deGasperin2017}. The study of these sources as a population sheds light upon the long-standing problem of the presence and properties of a seed population of cosmic-ray electrons (CRe) in the diffuse intra-cluster medium. The existence of such a population would mitigate the limitation of some standard cosmic-ray acceleration theories such as the diffusive shock acceleration (DSA) of thermal pool electrons \citep{Kang2014}.

\subsection{Radio-loud AGN}

\lol{} will provide the lowest frequency data-points for a large variety of radio AGN spectra, ranging from young (few hundred years) gigahertz-peaked spectrum and compact steep spectrum radio sources to old ($\sim 10^8$ years) giant Mpc-sized radio galaxies \citep[e.g.][]{Shulevski2019, Dabhade2019}. In compact objects, this information can be used to distinguish between jets which may be `frustrated' and not powerful enough to clear the medium and propagate outside the host galaxy or a `young' scenario, in which the radio AGN may only recently have become active (see e.g. \citealt{Callingham2017} and \citealt{ODea2020} for a recent review). By measuring the properties of the low-frequency turnover in compact sources and hotspots, we can evaluate the relative importance of synchrotron self-absorption, free-free absorption, or a low-energy cut-off \citep[e.g.][]{McKean2016, deGasperin2020}. Furthermore, the combination of \lol{}, LoTSS and higher-frequency surveys such as NVSS or Apertif will enable the study of spectral curvature over a wide frequency band for a large number of sources ($\sim1,000,000$; considering the northern hemisphere). Such a statistical sample can be used to characterise the overall shape of the radio spectral energy distribution (SED) and examine how it changes with stellar mass and redshift. Spatially resolved spectral studies combining \lol{} and LoTSS will be possible for samples of thousands of nearby or physically large AGN \citep[e.g. 1500 radio galaxies with size $> 60\arcsec$ in the HETDEX region;][]{Hardcastle2018}. The LBA survey data could also shed some light on the activity cycles in the newly-discovered population of low-luminosity FRIIs \citep{Mingo2019}, which are believed to inhabit lower mass hosts than their high luminosity counterparts.

In the case of blazars (radio-loud AGN whose relativistically beamed jets are oriented close to the line of sight), radio-spectral indices are characteristically flat throughout the centimetre band and even down to the LOFAR HBA band at $\sim 150$\,MHz \citep{Trustedt2014,Mooney2019}. These flat spectra are due to the superposition of many different jet emission zones near the compact base of the jets of varying size and synchrotron turnover frequency. At sufficiently low frequencies, however, the blazar emission is expected to become altered by strong self-absorption. On the other hand an additional component of unbeamed steep-spectrum emission from the optically extended jets and lobes might start to dominate the source emission. Due to their flat spectra, blazars are generally much fainter at MHz frequencies than unbeamed radio-loud AGN so that their properties in this regime are poorly studied. \citet{Massaro2013, Giroletti2016, DAntonio2019} found that the average radio spectrum of large samples of blazars is flat down to tens of MHz, suggesting that their spectra are still dominated by the beamed core emission even at such ultra-low frequencies. However, previous studies were affected by variability and limited angular resolution, which rendered it impossible to separate the core and lobe emission of blazars. This will be improved significantly by \lol{} and follow-up LOFAR LBA observations.

Blazars are also an important source class for high-energy astronomy and astroparticle physics. \citet{Mooney2019} found low-frequency radio counterparts to all gamma-ray sources in the Fermi Large Area Telescope Third Source Catalog \citep[3FGL][]{Acero2015} at 150\,MHz within LoTSS that are associated with known sources at other wavelengths and found source candidates for unassociated gamma-ray sources within the LoTSS footprint. Covering the same field, \lol{} opens the opportunity to unveil (possibly new) associations at even lower frequencies.

Ultra-low frequency data are also crucial for the study of remnants of radio galaxies \citep{Brienza2017, Mahatma2018}. Recent investigations have shown that this elusive population of sources exhibits a range of spectral properties and some still show the presence of a faint core \citep{Morganti2020, Jurlin2020}. The addition of a very low frequency point for a resolved spectral analysis will constrain the time scale of the `off' phase. The ultimate aim is to obtain a census of AGN remnants that will provide the rate and duration of the AGN radio-loud phase, allowing for a comprehensive study of triggering and quenching mechanisms and constraining models of the radio activity in relation to the inter-stellar medium (ISM) and associated star formation rates. A key contribution to the quantitative study of the AGN life cycle will also come from the study of restarted radio galaxies whose identification and temporal evolution due to plasma ageing will also be possible only through the measurement of their low-frequency spectra \citep{Jurlin2020}. 

\lol{} and LoTSS data, combined with optical, infrared (IR), and millimetre data sets, will also be used to determine the evolution of black hole accretion over cosmic time and to address crucial questions related to the nature of the different accretion processes, the role of AGN feedback in galaxy evolution, and its relation to the environment. Dramatic examples of such feedback include the giant X-ray cavities seen in the hot atmospheres of many cool-core galaxy groups and clusters. These cavities, inflated by the lobes of the central AGN, represent an enormous injection of feedback energy. The low-frequency data of these systems are critical for constraining the state of the plasma in the largest cavities, as well as in later phases, when the relativistic plasma is effectively mixed by instabilities with the thermal ICM. In fact, old (`ghost') cavities from earlier generations of activity are often only visible at very low frequencies, due to ageing effects and large angular scales \citep[e.g.][]{Birzan2008}. These measurements will clarify the AGN duty cycle and the impact of AGN feedback by refining scaling relations between the radio and feedback power \citep[see][for a review]{Heckman2014}. Lastly, \lol{} will have the unique potential to reveal possible reservoirs of very old CRe that could explain the often observed discrepancy between the young spectral age of radio galaxies and the apparently older dynamical age \citep{Heesen2018, Mahatma2020}.

\subsection{Galaxies}

\lol{} will give access to the lowest radio frequencies in galaxies, so that we can study the radio continuum spectrum in unprecedented detail. The main science drivers are: (i) using the radio continuum as an extinction-free star formation tracer in galaxies; (ii) characterising radio haloes as a mean of studying galactic winds; and (iii) investigating the origin and regulation of galactic magnetic fields. 

Radio continuum emission in galaxies results from two distinct processes: thermal (free--free) and non-thermal (synchrotron) radiation. Both are related to the presence of massive stars, with UV radiation ionising the gas leading to free-free emission. The same stars end their lives in supernovae, which are the most likely places for the acceleration of CRe to GeV-energies, which are responsible for the synchrotron emission. 

The relationship between the radio continuum emission of a galaxy and its star formation rate (SFR), that is, the radio-SFR relation, is centred on the interplay of star formation and gas, magnetic fields, and CRe \citep{Tabatabaei2017}. At frequencies below 1 GHz the thermal contribution is less than ten percent for the global spectra,  which means that with low frequencies, we can study the non-thermal radio--SFR relation, which has more complex underlying physics. This is particularly the case when galaxies are not electron calorimeters, meaning that some CRe escape via diffusion and advection in winds. Hence, to make it possible to exploit the radio--SFR relation for distant galaxies at this frequency, we need to calibrate this relation in nearby galaxies with known SFRs \citep[e.g.][]{CalistroRivera2017}. As a side-effect, we can explore the physical foundation that gives rise to the relation in the first place, such as the link between magnetic field strength and gas density \citep[e.g.][]{Niklas1997}. Eventually, \lol{} will detect thousands of galaxies at $z < 0.1,$ providing data that can be used to distinguish between various models for the scarcely explored ultra low-frequency radio-SFR relation and its close corollary, the radio-far-infrared (radio-FIR) correlation, down to the frequencies where it may break down due to free--free absorption. These data will also explore the variation with galaxy properties \citep[as done at HBA frequencies by][]{Gurkan2018, Smith2020}, which are essential to constrain if radio data are to be used to probe star formation at higher redshifts.

Low-frequency observations are particularly useful for spatially resolved studies of CRe and magnetic fields in nearby galaxies. The distribution of the radio continuum emission is smoothed with respect to the CRe injection sites near star-forming regions. This can be ascribed to the effects of CRe diffusion, a view that is backed up if we use the radio spectral index as a proxy for the CRe age \citep{Heesen2018}. However, radio continuum spectra are also shaped by CRe injection, losses, and transport -- for instance, in the case of advection in galactic winds \citep[e.g.][]{Mulcahy2014}. Hence, a fully sampled radio spectrum from the MHz to the GHz regime is necessary for reliably assessing the age of CRe and also for disentangling the effect of free--free absorption. \lol{} data make it possible to detect the turnover from free-free emission with fairly low emission measures, probing low-density (5 cm$^{-3}$) warm ionised gas which may be prevalent in the mid-plane of galaxies \citep{Mezger1978}. Even though one particular statistical study using LOFAR HBA at 144~MHz hinted that free--free absorption plays a minor role \citep{Chyzy2018}, the contribution from cooler ($T<1000$~K) ionised gas remains largely uncertain \citep{Israel1990, Emig2020}. \lol{} will probe a critical turnover frequency in SED modelling that can characterise ionised gas properties and distinguish its contributions from CRe propagation effects. Furthermore, we are able to explore, for the first time, a possible deviation from a power-law cosmic-ray injection spectrum at the lowest energies.

\lol{} radio-continuum observations could open up a new avenue for studying galactic winds and their relation with the circum-galactic medium \citep[CGM; see][for a review]{Tumlinson2017}. Edge-on galaxies show extensive radio haloes, indicating the presence of CRe and magnetic fields. By enabling an analysis of the vertical spectral index profile, \lol{} data can be used to estimate the spectral age of the CRe and, thus, to measure the outflow speed of the wind. LOFAR has allowed some progress to be made, with radio haloes now detected to much larger distances than what was previously possible \citep{Miskolczi2019}. The LOFAR LBA system is likely to detect radio haloes to even grater distance, thereby providing deeper insights on galaxies interaction with the CGM.

Finally, \lol{} could lead to fundamental constraints on the nature of dark matter in dwarf spheroidal galaxies. One of the leading candidates for this are weakly interacting massive particles (WIMPs), which can produce a radio continuum signal annihilating electron-positron pairs. For typical magnetic field strengths, the peak of the signal is expected in the hundred megahertz frequency range if the WIMPs are in the mass range of a few GeV. The LOFAR HBA search by \cite{Vollmann2020} has so far provided upper limits, which can possibly improved with LOFAR LBA observations, particularly in the lower mass range where HBA observations are less sensitive.

\subsection{The Milky Way}

Low-frequency observations with LOFAR will open a new area of discovery space in Galactic science. \lol{} will image a large fraction of the northern Galactic Plane, thereby completing a census of supernova remnants (SNR). This will enable a search for the long-predicted and missing population of the oldest SNR, whose strongly rising low frequency spectra and large angular scales are not visible at higher frequencies \citep{Driessen2018, Hurley-Walker2019}. The combination of \lol{} and LoTSS data will be important in identifying emission from \Hii{} regions whose morphology is similar to that of SNRs, whilst having a flatter spectrum. Additionally, \lol{} data will enable the measure the low-frequency spectral curvature of supernova remnants as a diagnostic of shock acceleration and the foreground free-free absorption \citep{Arias2018}. 

\lol{} will provide a map of Galactic non-thermal emission \citep[e.g.][]{Su2017} and it can also map and characterise the properties of self absorption by low-density ionised gas that appears as `absorption holes' against the smoother background. Concurrently, such observations will serve as a proxy to tomographically image the CRe distribution and magnetic field configuration throughout the Galaxy \citep{Polderman2019, Polderman2020}. Finally, \lol{} will enable the study of: (i) the role of pulsar wind nebulae in dynamically shaping their environment; (ii) the star-forming processes in close proximity to very young stellar objects by detecting their associated thermal and non-thermal emitting radio jets; and (iii) candidate pulsars through their ultra-steep spectra. 

\subsection{Stars and Exoplanets}

Radio emission from stars is a key indicator of magnetic activity and star-planet plasma interactions \citep{Hess2011}. Existing studies have mainly focused on cm wavelengths ($\nu>$ 1 GHz), and have been largely restricted to a small subset of anomalously active stars such as flare stars \citep[e.g. UV Ceti;][]{Lynch2017} and close binaries \citep[e.g. colliding wind-binaries;][]{Callingham2019}. Recently,  the first metre-wave (120 -- 170 MHz) detection of a quiescent M-dwarf, GJ 1151, was taken by LoTSS, with flux density of 0.8 mJy and $>$60\% circular polarisation \citep{Vedantham2020}.The emission characteristics and stellar properties strongly suggest that the low-frequency emission is driven by a star-exoplanet interaction. In parallel, weak radio bursts from the Tau Bootes system that hosts a hot Jupiter have been tentatively detected in the $14-21$ MHz range using LOFAR in beamformed mode \citep{Turner2020}. These discoveries herald an unprecedented opportunity to constrain magnetic activity in main-sequence stars other than the Sun as well as the impact of the ensuing space-weather on exoplanets, as exemplified by the 19 other detections presented by Callingham et al. (under review). Additionally, the recent direct discovery of a cold brown dwarf using LoTSS data \citep{Vedantham2020a} also demonstrates the new potential of deep low-frequency surveys in helping us to understand the properties of planetary-scale magnetic fields outside of the Solar System. 

Since the detected radio emission is produced via the electron cyclotron maser instability (ECMI), the frequency of emission is directly related to magnetic field strength of either the star or exoplanet. Therefore, at HBA frequencies, studies are restricted to a subset of extreme M and ultracool dwarfs that have strong magnetic fields ($> 50$ G). With its lower frequencies, \lol{} can begin to probe exoplanets and stars with magnetic field strengths similar to those found in our Solar System ($\sim 5$ to 50 G), implying that we should be sensitive to a Jupiter-Io-like system out to 10 pc. The most stringent upper limit on such a detection has been carried out using LOFAR LBA data \citep{deGasperin2020b}. If the discovery rates and luminosities of the systems stay similar to those derived from LoTSS, we would expect $35\pm15$ detections in the complete \lol{} (Callingham et al. under review). Hence, \lol{} will play a major role in characterising the phenomenology of low-frequency emission of stellar systems and has the potential to dramatically impact on our understanding of the magnetic field properties and environments of other planetary systems around nearby stars. 

\subsection{Ionosphere}

Continuous, systematic, long-term observations at very low-frequencies will allow for the characterisation of important aspects of the ionosphere, such as physical parameters of ionospheric travelling waves, scintillations, and the relation with solar cycles \citep{Mevius2016, Helmboldt2020}. All of these are crucial aspects for constraining ionospheric models. Instruments observing at ultra-low frequency are powerful tools for deriving the total electron content (TEC) of the ionosphere independently from standard observations with satellite measurements \citep{Lenc2017, deGasperin2018a}. \lol{} observations will also provide large data sets with which it is possible to study the higher-order effects imprinted on travelling radio-waves as Faraday rotation and the ionospheric third-order delay \citep{deGasperin2018a}.

\subsection{The unusual and unexpected}

Serendipitous discoveries have always played an important role in astronomy, particularly with the opening of new spectral windows. An example is the transient detected during the initial years of LOFAR observations, whose nature is still unclear \citep{Stewart2016}. \lol{} probes the lowest energy extreme of the electromagnetic spectrum, a regime where exotic radiation mechanisms such as plasma oscillations play a role. A potentially exciting part of analysing \lol{} will be searching for new, unexpected classes of objects that are only detectable at or below 50 MHz.  

\section{The LOFAR LBA sky survey}
\label{sec:lolss}

Building on the performance statistics of LOFAR during commissioning observations, we selected an observing mode for \lol{} that optimises survey speed whilst achieving the desired angular resolutions of 15\arcsec{} and sensitivity of $\sim 1$~\mjybeam. A summary of the final observational setup is listed in Table~\ref{tab:specs}.

\begin{table}
\centering
\begin{threeparttable}
\begin{tabular}{lc}
\hline\hline
Number of pointings & 3170 \\
Separation of pointings & 2.58\deg \\
Integration time (per pointing) & 8 h \\
Frequency range & 42--66 MHz \\
Array configuration & LBA OUTER \\
Angular resolution & $\sim15$\arcsec \\
Sensitivity & $\sim 1$~\mjybeam \\
Time resolution & 1 s \\
-- after averaging & 2 s \\
Frequency resolution & 3.052 kHz \\
-- after averaging & 48.828 kHz \\
\hline
\end{tabular}
\end{threeparttable}
\caption{\lol{} observational setup.}\label{tab:specs}
\end{table}

The LOFAR LBA system has the capability of simultaneously casting multiple beams in different and arbitrary directions at the expense of reduced observing bandwidth. In order to maximise the survey speed and to provide an efficient calibration strategy, during each observation we continuously keep one beam on a calibrator source whilst placing three other beams on three well-separated target fields. The beam on the calibrator is used to correct instrumental (direction-independent) effects such as clock delays and bandpass shape \citep{deGasperin2019}; the rationale behind continuously observing the calibrator is that these systematic effects are not constant in time and can be more easily derived from analysing well-characterised calibrator fields.

Ionospheric-induced phase variations are the most problematic systematic effect at ultra-low frequencies \citep{Intema2009, Mangum2014, Vedantham2014, Mevius2016, deGasperin2018a}. In order to mitigate the consequences of poor ionospheric conditions on a particular observation, we used the following observing strategy: during each observation we simultaneously place three beams on three target fields for one hour. After an hour, we switch beam locations to a different set of three targets. We schedule observations in 8 hour blocks. In total, 24 fields are observed for one hour each in an 8 hour block. The same process is then repeated eight times to improve the sensitivity and the $uv$-coverage of each pointing. In this way, if the ionosphere was particularly problematic during a particular day, it would have affected only a fraction of the data in each field, without compromising the uniform sensitivity of the coverage. To prepare the observations, we use a scheduling code that implements this observing strategy, whilst maximising the $uv$-coverage so that each field is not observed twice at hour angles closer than 0.5 hr. We ensure that observations were taken when the Sun is at least 30\deg{} from the targeted fields and their elevation is above 30\deg{}.

The total bandwidth available in a single LOFAR observation is 96 MHz (8-bit mode). When divided into four beams this gives a usable band of 24 MHz. We tuned the frequency coverage to 42--66 MHz to overlap with the most sensitive region of the LBA band taking into account both the sky temperature and the dipole bandpass \citep[see][]{VanHaarlem2013}. To suppress the effect of strong radio frequency interference (RFI) reflected by ionospheric layers at frequencies $<20$ MHz, the LBA signal path is taken through a 30-MHz high-pass filter as default.

Due to a hardware limitation (which will be removed in a future upgrade to LOFAR) in each station, only half of the LBA dipoles can be used during a single observation. The choice of the dipoles that are used has a large impact on the size and shape of the main lobe of the primary beam and on the positions and amplitudes of the side lobes. The LOFAR LBA system can be used in four observing modes:

\begin{description}
  \item[LBA INNER:] The inner 48 dipoles of the station are used. This mode gives the largest beam size at the cost of a reduced sensitivity. The calibration of the inner dipoles (the station calibration) is less effective than for the outer dipoles due to mutual coupling and their higher sensitivity to Galactic emission during the station calibration procedure. The effective size of the station is 32 m, which corresponds to a primary beam full width at half maximum (FWHM) of 10\deg{} at 54 MHz.
  \item[LBA OUTER:] The outer 48 dipoles of the station are used. This mode minimises the coupling between dipoles but reduces the beam size. The effective size of the station is 84 m, providing a primary beam FWHM of 3.8\deg{} at 54 MHz.
 \item[LBA SPARSE (ODD or EVEN):] Half of the dipoles, distributed across the station, are used. At the time of writing this mode is experimental, but grants an intermediate performance between LBA INNER and OUTER, with a suppression of the magnitude of the side-lobes compared to the latter. The effective size of the station is around 65 m, which provides a primary beam FWHM of 4.9\deg{} at 54 MHz.
\end{description}
Given the better quality of the LBA OUTER station calibration and the close similarity of the primary beam FWHM with the HBA counterpart (3.96\deg{} at 144 MHz) this observing mode was used. The LBA OUTER mode results in a primary beam FWHM ranging from 4.8\deg{} to 3.1\deg{} for the covered frequency range between 42--66 MHz. The use of the LBA OUTER mode also implies the presence of a non-negligible amount of flux density spilling in from the first side lobe. This effect is partially compensated for by the calibration strategy, where sources in the first side lobe are imaged and subtracted \citep[see][]{deGasperin2020a}.

\begin{figure*}
\centering
 \includegraphics[width=\textwidth]{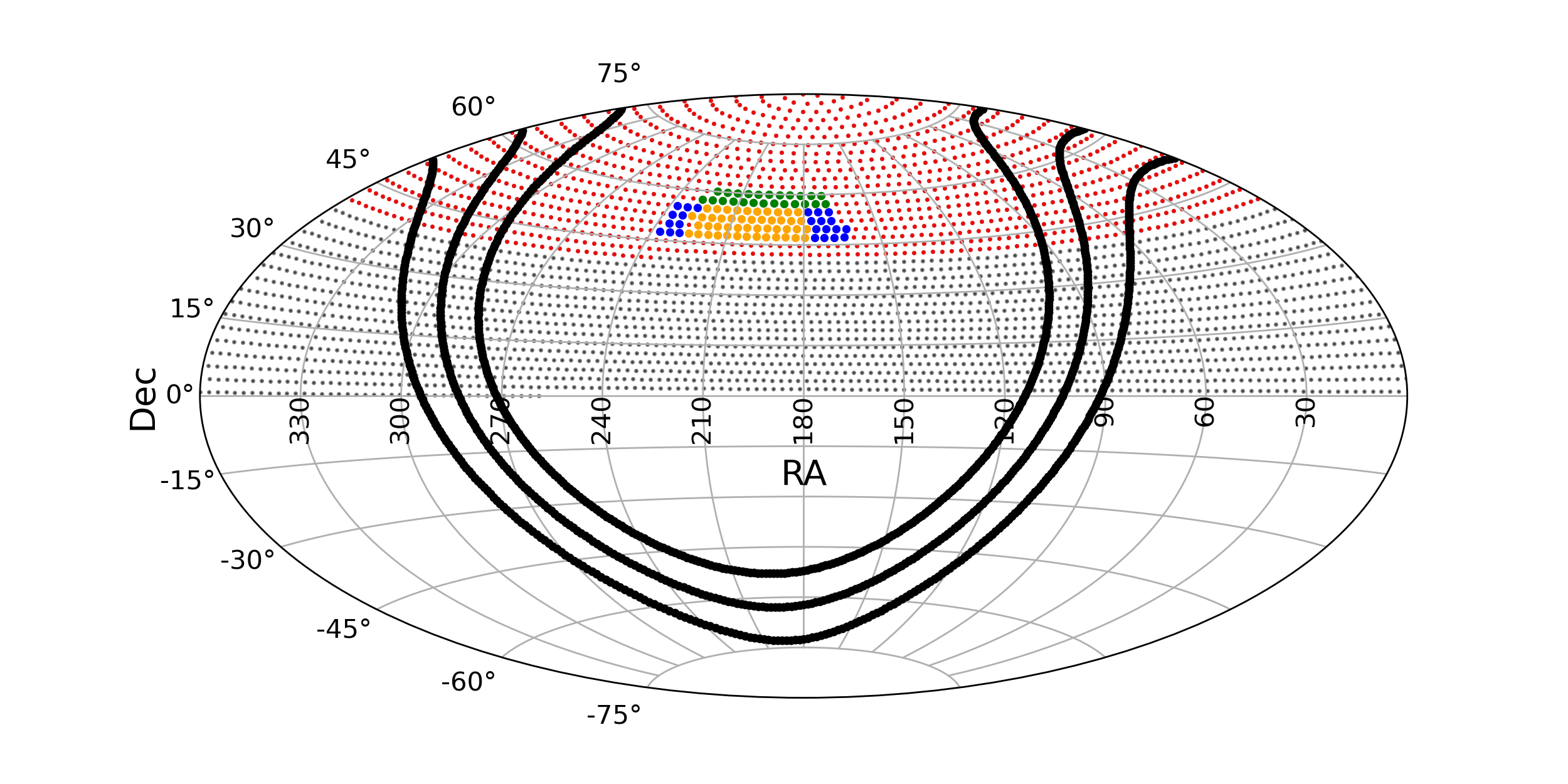}
 \caption{Current and planned sky coverage of \lol{}. Each dot is a pointing of the full survey. Red dots are scheduled to be observed by 2022 with a priority on extragalactic fields. The region presented in this paper is coloured in yellow (cycle 8 data), blue (cycle 9 data) and green (cycle 12 data). Solid black lines show the position of the Galactic plane with Galactic latitude: $-10$\deg, 0\deg, $+10$\deg.}
 \label{fig:coverage}
\end{figure*}

Since the FWHM of the primary beam of LoTSS and \lol{} is similar, we adopted a joint pointing strategy so that each target field is centred on the same coordinates in both surveys. \lol{} therefore has the same pointing scheme as LoTSS (see Fig.~\ref{fig:coverage}). The pointing scheme follows a spiral pattern starting from the north celestial pole, with positions determined using the \cite{Saff1997} algorithm. This algorithm attempts to uniformly distribute points over the surface of a sphere when there is a large number of pointings. Using the same pointing separation as LoTSS (2.58\deg), the coverage of the entire northern hemisphere requires 3170 pointings. Assuming circular beams, this separation provides a pointing distance of FWHM/1.2 at the highest survey frequencies and better than FWHM/$\sqrt{3}$ at the lowest. The distance between pointings at the mean frequency is close to FWHM/$\sqrt{2}$.

Currently, LOFAR is composed of 24 core stations (CS), 14 remote stations (RS), and 14 international stations (IS). The CS are spread across a region of radius $\sim2$~km and provide 276 short baselines. The RS are located within 70~km from the core and provide a resolution of $\sim15\arcsec$ at 54~MHz, with a longest baseline of 120~km. \lol{} makes use of CS and RS, whilst IS data were not recorded to keep the size of the data set manageable\footnote{The use of IS would have increased the data set size by a factor of $\sim10$. A factor of 2 of this would come from more baselines, and a factor of 4 to 8 from the increase in the frequency-time resolution required in order to account for larger differential ionosphere on the longest baselines.}. For an example of the $uv$-coverage, which by design can be different for each pointing, we refer to \cite{deGasperin2020a}. The longest baseline available for the observations presented in this paper was approximately 100 km, providing a nominal resolution at mid-band (54 MHz) of 15\arcsec.

The final aim of \lol{} is to cover the northern sky to a depth of $\sim 1$~\mjybeam. With the LBA system, this requires around 8 hrs of integration time at optimal declination, although the final noise is mostly limited by ionospheric conditions and experiments indicate that in practice, it will range between 1 and 1.5~\mjybeam \citep{deGasperin2020a}. In this preliminary release, where the direction-dependent errors are not corrected, the noise ranges between 4 and 5~\mjybeam.

Ionospheric scintillations can make ultra low-frequency observations challenging by decorrelating the signal even on very short baselines. Several years of observations of the amplitudes of ionospheric scintillations using LOFAR show that the phenomenon is more prevalent from sunset to midnight than during the daytime (priv. comm. R. Fallows), which broadly follows patterns that have been observed at higher latitudes \citep{Sreeja2014}. Therefore, in order to minimise the chances of ionospheric scintillations, all the observations presented here were taken during daytime. However, daytime observations have some drawbacks, predominantly at the low-frequency end of the full LBA band (i.e. $<30-40$~MHz), below the frequency coverage of \lol{}. Due to solar-induced ionisation, the ionosphere becomes thicker during the day. This has two main consequences: the lower ionospheric layers can reflect man-made RFI towards the ground, which is typically seen at frequencies $<20$~MHz. At the same time, the effect of Faraday rotation is expected to be larger, because it also depends on the absolute total electron content of the ionosphere, which can increase by a factor of 10 in the daytime. Since Faraday rotation has a frequency dependency of $\nu^2$, this systematic effect is dominant at the lowest-frequency end of our band coverage, where the differential rotation angle on the longest baselines is typically one to two radians \citep{deGasperin2019}.  

The resolution in time and frequency is chosen to minimise the effect of time and frequency smearing at the edge of the field of view as well as to be able to track typical ionospheric variations, whilst avoiding  the compilation of data sets that are too large to handle. Data are initially recorded at 1 s / 3.052 kHz resolution and are then flagged for RFI \citep{Offringa2010a} and bright sources removed from far side lobes \citep{deGasperin2019}. Before they are stored into the LOFAR Long Term Archive \footnote{\url{https://lta.lofar.eu/}}, the data are further averaged to 2 sec and 48.828 kHz\footnote{This corresponds to 4 channels per Sub Band (SB), where a SB bandwidth is 195.3125~kHz wide.}. 

The effects of the time and bandwidth smearing due to this averaging can be approximated using the equations of \cite{Bridle1989}. At a distance of 2\deg{} from the phase centre and at 15\arcsec{} resolution, the time averaging to 2 sec leads to a time smearing that reduces the peak brightness of sources by $<1\%$. At the same distance from the phase centre and resolution, frequency averaging to 48.828 kHz causes a frequency smearing that reduces the peak brightness of sources by about $7\%$. The time averaging period is kept short to allow for the calibration process to track rapid ionospheric variations and thus avoid decorrelation and the subsequent loss of signal.
Within the chosen frequency resolution of 48.828 kHz, a differential (between stations) TEC value of 1 TEC unit (TECU; $10^{16}$ electrons m$^{-2}$) produces a phase variation of 13\deg{} at 42 MHz. Typical variations within LOFAR core and remote stations are well within 1 TECU and can therefore be corrected in each channel without signal loss. The highest differential variation that can be tracked within a 2 sec time slot is about $10$ mTECU, corresponding to a drift in phase of $\sim 115\deg$ at 42 MHz.

\section{Survey status}
\label{sec:survey_status}

Using the survey strategy described above, the full northern sky can be observed in $3170\ {\rm pointings}\ /\ 3\ {\rm beams}\ \times\ 8\ {\rm hours}\ =\ 8453\ {\rm hours}$, although low-declination observations are still experimental. We have collected an average of 8 hrs of data on 95 pointings (3\% of the coverage), which covers, at full depth, about 500 deg$^2$ (see Fig.~\ref{fig:coverage}). These observations are concentrated around the HETDEX spring field\footnote{RA: 11 h to 16 h and Dec: 45\deg{} to 62\deg{} in the region of the Hobby-Eberly Telescope Dark Energy Experiment (HETDEX) Spring Field \citep{Hill2008}.} and are the focus of this preliminary data release. Of the 95 pointings, 19 have only 7 hrs of usable observation due to various technical problems. One field (P218+55) has been observed for 16 hrs and another field is currently missing from the coverage (P227+50) but will be added to the survey in a future release.

Archived data includes full Stokes visibilities from all Dutch stations (core and remote) but not from the international stations. The frequency coverage is always 42 -- 66 MHz. The data are also compressed using the Dysco algorithm \citep{Offringa2016}. Archived data are already pre-processed to flag RFI before averaging and to subtract the effect of Cygnus A and Cassiopeia A if some of their radiation was leaking through a far side lobe. The data size for an observation of 8 hrs is $\sim 100$ GB per pointing.

The present allocated observing time allows for the coverage of all fields above 40\deg{} declination. This campaign will cover 6700 deg$^2$ (1035 pointings), that is 33\% of the northern sky with 3 hrs per pointing, reaching a sensitivity of 2~\mjybeam. Lower declination and full depth are planned for future observing campaigns.

\subsection{Ionospheric conditions}
\label{sec:ionosphere}

The LOFAR LBA data have been collected since cycle 0 (2013), which was close to the solar maximum. This led to rapid and strong variations in the ionospheric properties in the years between 2011-2014, posing a particular challenge for  data processing. From 2014 onwards,  solar activity steadily decreased to reach a minimum in 2020. The quality of the LBA data steadily increased with decreasing solar activity and we achieved close to 100\% usable data in cycle 8 (2017). Currently,  solar activity is close to its minimum, which for this solar cycle had been particularly long \citep{DeToma2010}. If the next solar cycle is similar to the past one as predicted, the good conditions for low frequency observations will continue until around 2022. Solar activity will then make low-frequency observations challenging until 2027.

\section{Data reduction}
\label{sec:reduction}

The data reduction of \lol{} is being carried out in a distributed manner on computing clusters located at the Observatory of Hamburg, the Observatory of Leiden, the Institute of Radio Astronomy (INAF, Bologna), and the University of Hertfordshire. Synchronisation between the various running jobs is maintained through a centralised database. All computations are carried out in the same environment built within a Singularity container based on Ubuntu 20.04.

Here we present a preliminary release of \lol{} data, which was prepared using the automated Pipeline for LOFAR LBA (PiLL)\footnote{The pipeline for the data reduction is publicly available at \url{https://github.com/revoltek/LiLF}}, that is described in detail in \citet[][for the calibrator processing]{deGasperin2019} and in \citet[][for the target processing]{deGasperin2020}. PiLL now includes the possibility of carrying out full direction-dependent calibration. However, because this is still experimental, in this paper we discuss and release only those data sets that have been processed as far as the direction-independent calibration. A direction-dependent survey release reaching 1~\mjybeam{} root mean square (rms) noise and 15\arcsec{} resolution will be presented in a forthcoming publication.

\begin{table}
\centering
\begin{threeparttable}
\begin{tabular}{lccc}
Cycle & Proposal  & Year & N pointings \\
\hline
8     & LC8\_031  & 2017 & 47    \\
9     & LC9\_016  & 2017 & 24    \\
12    & LC12\_017 & 2019 & 24    \\
\hline
\end{tabular}
\end{threeparttable}
\caption{Observations used for the preliminary release.}\label{tab:obs}
\end{table}

\subsection{Calibration}
\label{sec:calibration}

Depending on the target position, the scheduler code selects the closest calibrator from 3C\,196, 3C\,295, and 3C\,380. Here, we summarise the most important calibration steps. Following \cite{deGasperin2019}, the calibrator data are used to extract the polarisation alignment, the Faraday rotation (in the direction of the calibrator), the bandpass of each polarisation and phase solutions, which include the effects of both the clock and ionospheric delay (in the direction of the calibrator). For each hour of observation, the polarisation alignment, bandpass, and phase solutions are then transferred to the three simultaneously observed target fields. Direction-independent effects are then removed to produce target phases that now include a differential ionospheric delay with respect to the calibrator direction.

All eight data sets for each target field, which have generally been observed on different days, are then combined prior to performing the self-calibration procedure. The initial model for the self calibration is taken from the combination of TGSS \citep{Intema2017}, NVSS \citep{Condon1998}, WENSS \citep{Rengelink1997}, and VLSSr \citep{Lane2014} and includes a spectral index estimation up to the second order that is used to extrapolate the flux density to \lol{} frequencies. As explained in \cite{deGasperin2020}, the first systematic effect to be calibrated is the ionospheric delay, followed by Faraday rotation and second order dipole-beam errors. The latter is the only amplitude correction. Since it is a correction on the dipole beam shape, it is constrained to be equal for all stations. It reaches a maximum of few per cent level and it is normalised to ensure that the flux density scale is not altered by our initial self-calibration model. Finally, sources in the first side lobe are imaged and removed from the data set. The process is repeated twice to improve the model and the output is a direction-independent calibrated image.

\subsection{Imaging}
\label{sec:imaging}

Final imaging is performed with WSClean \citep{Offringa2014} with Briggs weighting $-0.3$ and multi-scale cleaning \citep{Offringa2017}. The maximum $uv$ length used for the imaging is 4500$\lambda$, which removes the longest baselines that are more seriously impacted by large direction-dependent ionospheric errors. A third order polynomial is used to regularise the spectral shape of detected clean components. The gridding process is performed using the Image Domain Gridder algorithm \citep[IDG;][]{VanDerTol2018}. The implementation of IDG in WSClean allows for the imaging of visibilities whilst correcting for time and direction-variable beam effects.

\begin{figure*}[htb]
\centering
 \includegraphics[width=\textwidth]{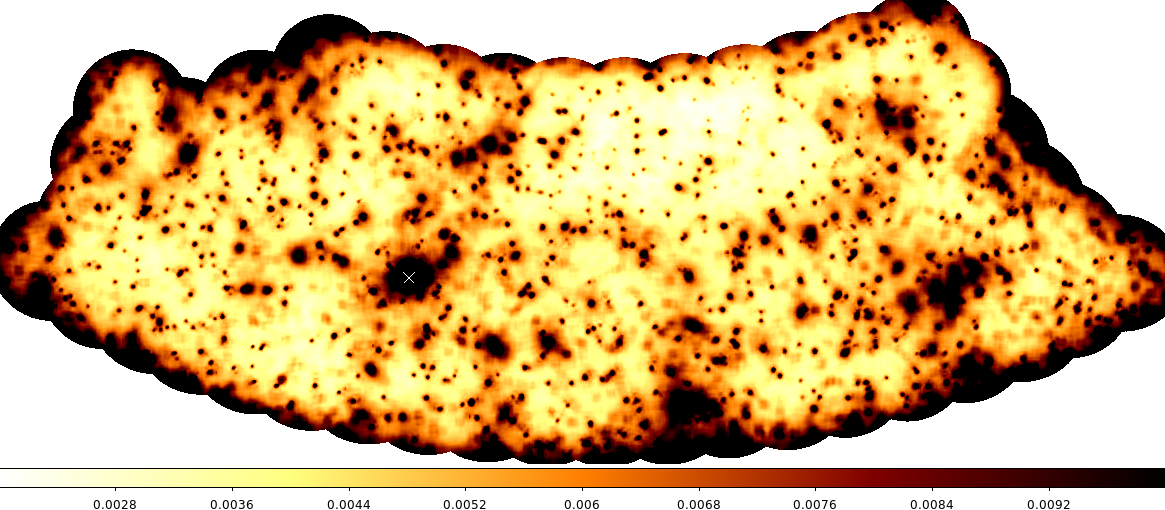}
 \caption{Rms noise map of the HETDEX region in Jy beam$^{-1}$. The regions with reduced sensitivity are located at the edges of the survey footprint and around bright sources. The location of 3C\,295 is marked with a white X, the presence of the bright source increases the local rms substantially.}
 \label{fig:rms}
\end{figure*}

The 95 direction-independent calibrated images are combined into a single large mosaic. Sources from each image are cross-matched with sources from the FIRST catalogue \citep{Becker1995} to correct the astrometry for each pointing independently (see also Sect.~\ref{sec:astrometry}, for a full description of the process). Only 'isolated' and 'compact' sources are used for this cross-matching process. In order to be considered as isolated, a source needs its nearest neighbour to be at a distance of $>3\times47\arcsec$. To be defined as compact, a source needs to have its integrated to peak flux ratio lower than 1.2. The maximum shift applied to a field was of 2.9\arcsec, with the majority of the corrections being less than 1\arcsec. During the process all images are convolved to the minimum common circular beam of 47\arcsec. Pixels common to more than one pointing are averaged with weights derived by the local primary beam attenuation combined with the global noise of the pointing where each pixel belongs. All regions where the attenuation of the primary beam was below 0.3 were discarded during the process.

\section{Results}
\label{sec:results}

Here, we present the images from the preliminary data release, focusing on its sensitivity, astrometric precision, and accuracy, and assessing the uncertainty on the flux density. For the purpose of source extraction, we used PyBDSF \citep{Mohan2015}. The source extractions are made using a $4\sigma$ detection threshold on islands and a $5\sigma$ threshold on pixels. To reduce false positives, we used an adaptive rms box size that increases the background rms noise estimation around bright sources.

\subsection{Sensitivity}
\label{sec:sensitivity}

An image showing the local rms noise distribution calculated with PyBDSF is presented in Fig.~\ref{fig:rms}. The quality of the images varies significantly across the covered area, with regions with rms noise up to three times higher than others. The generally lower noise in the upper part of the region presented here is likely related to the observing period of the different fields. The upper region was observed during cycle 12 (2019) when the solar cycle was at its minimum, thus reducing the presence of ionospheric disturbances, whilst the southern part was observed during cycles 8 and 9 (2017).

\begin{figure}
\centering
 \includegraphics[width=.5\textwidth]{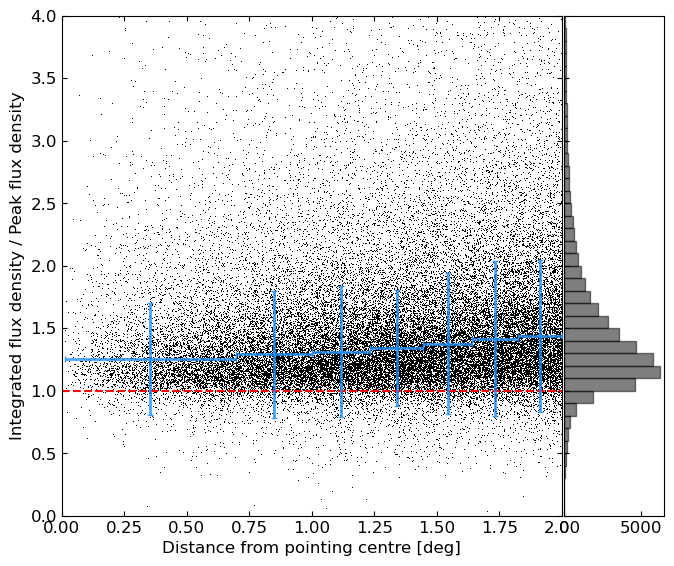}
 \caption{Ratio between the integrated flux density and the peak flux density of isolated sources, i.e. sources with no other detections closer than $3\times$ the beam size (47\arcsec), as a function of the distance from the pointing centre. Unresolved sources should have a value of around unity (red line), with resolved sources having higher values. The binned medians (blue crosses) go from 1.2 to 1.4. Since the majority of the sources in our catalogue is expected to be unresolved at this angular resolution, this is an indication of ionospheric smearing.}
 \label{fig:inttopeak}
\end{figure}

The ionospheric irregularities introduce phase errors that can move sources in a way that is not synchronised across the image and with a positional change that is non-negligible compared to the synthesised beam. Therefore, the main driver of the non-uniformity of the rms noise distribution is the presence of bright sources in combination with the limited dynamic range, caused by the time- and direction-varying ionosphere which cannot be corrected in a direction-independent calibration. Although we limited the length of the baselines to reduce this effect, the sources are still blurred and their peak flux is reduced (see Fig.~\ref{fig:inttopeak}). As expected, the effect is slightly more relevant for observations taken in 2017 (mean integrated-to-peak flux density ratio: 1.6) than for observations taken in 2019 (mean integrated-to-peak flux density ratio: 1.5), those that cover the northern region of the presented footprint. This effect and the non-uniformity of the rms noise will be reduced with the full direction-dependent calibration.
 
\begin{figure}
\centering
 \includegraphics[width=.5\textwidth]{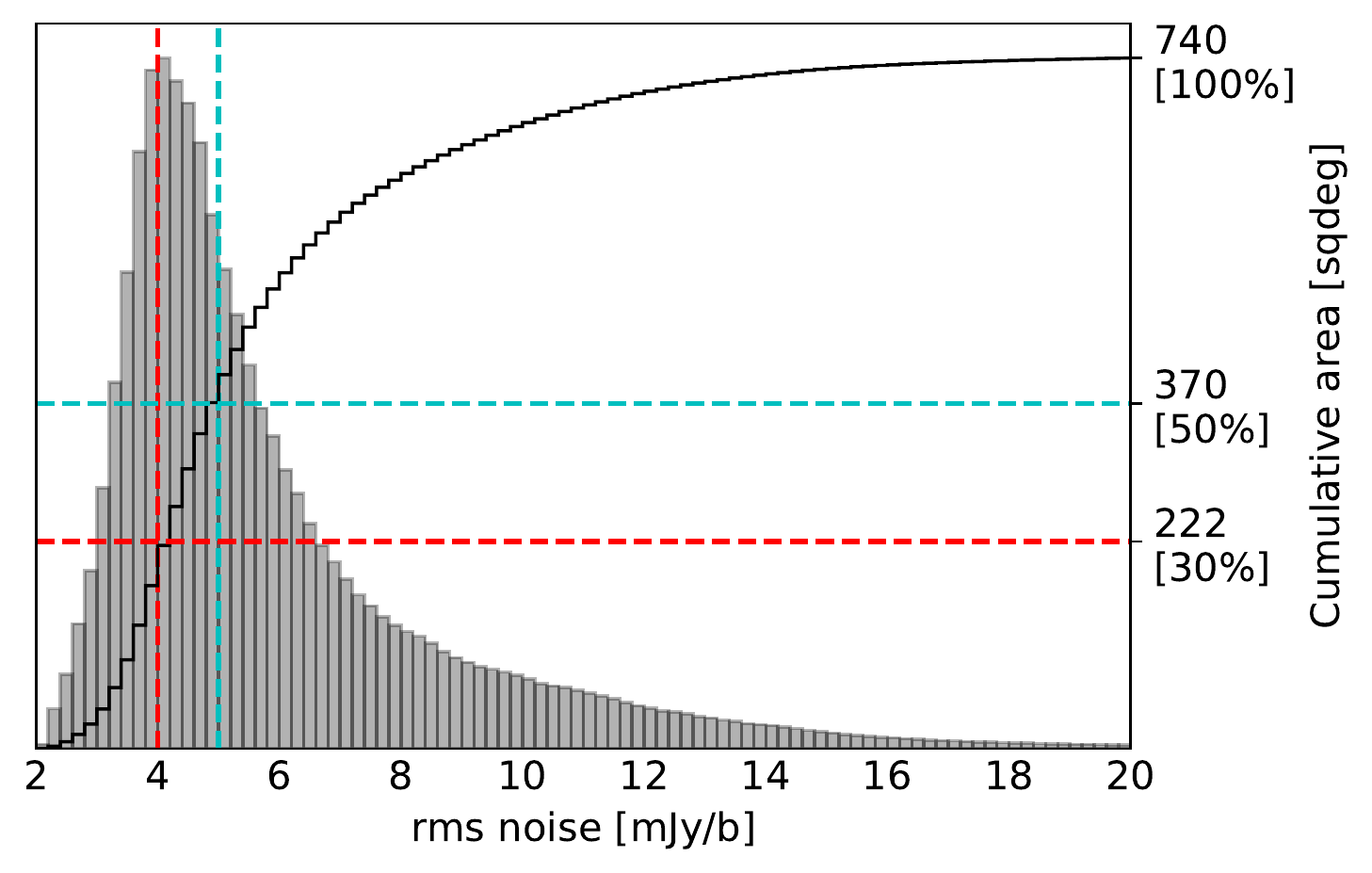}
 \caption{Histogram of the rms noise. The black solid line shows the cumulative function. The red dashed lines show that 30\% of the survey footprint (222 deg$^2$) has a rms noise $<4$~\mjybeam, whilst the blue dashed lines show that 50\% of the survey footprint (370 deg$^2$) has an rms noise $<5$~\mjybeam. The tail of noisy regions above 8~\mjybeam{} are due to the footprint edges and dynamic range limitations due to bright sources.} \label{fig:rmshist}
\end{figure}

In Fig.~\ref{fig:rmshist}, we show the histogram of the rms noise across the field. The histogram includes the edges of the field, where the noise is higher because of reduced coverage. Most of the covered region has a rms noise of $\sim 4$~\mjybeam. The area with a noise equal or lower than 4~\mjybeam{} is 222 deg$^2$ which accounts for 30\% of the presented region. The median rms noise of the entire region is $\sim 5$~\mjybeam. 

\subsection{Astrometric precision and accuracy}
\label{sec:astrometry}

The astrometric accuracy of our observations might be affected by errors in the initial sky model used for calibration, which was formed from a combination of catalogues from surveys at different resolutions. These errors might propagate through the phase solutions and introduce systematic errors in the position of our sources. However, the phase calibration is performed with a reduced number of degrees of freedom (one per antenna per time slot) thanks to the frequency constraint, which assumes that phase errors are largely due to TEC-induced delays. Systematic positional offsets are corrected during the mosaicing process (see Sect.~\ref{sec:calibration}). The way we identify positional offsets to correct during the mosaicing process and the way we assess our final accuracy are the same and we describe them in detail below.

\begin{figure}
\centering
 \includegraphics[width=.5\textwidth]{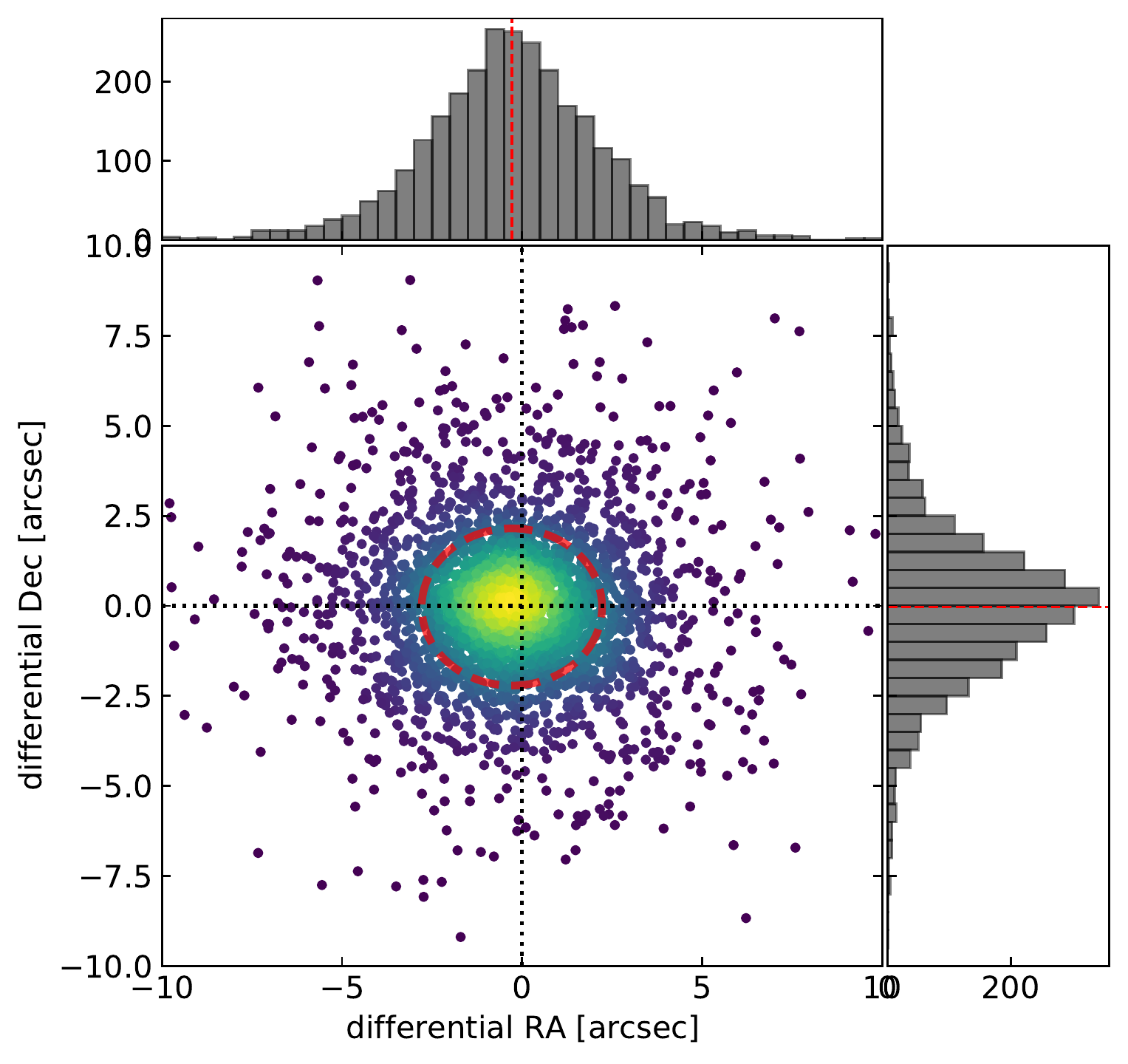}
 \caption{Astrometric accuracy of the sources in the catalogue (see text for the calculation). The average astrometric offsets are RA$=-0.28\arcsec$ and Dec$=-0.03\arcsec$ with relative standard deviations RA=$2.50\arcsec$ and Dec$=2.18\arcsec$. The red ellipse traces the standard deviation.
}
 \label{fig:astrometry}
\end{figure}

As a reference catalogue we used the FIRST survey, which has a systematic positional error of less than 0.1\arcsec{} from the absolute radio reference frame, which was derived from high-resolution observations of selected calibrators \citep{White1997}. To reduce the bias due to erroneous cross-matching we first reduce the \lol{} catalogue to isolated sources, namely those sources with no other detections closer than three times the beam size (47\arcsec). This brings the number of sources from 25247 to 22766 (90\%). This process was repeated for the reference catalogue, where only sources with no other detections within 47\arcsec{} were selected, reducing the number of sources by 35\%. This step is important to avoid selecting double sources, which are rather common. Then, the subset of sources of both \lol{} and the FIRST catalogues are further reduced to include only compact sources. Compact sources are defined as those with a total-flux to peak-flux ratio less than $1.2$. This reduces the \lol{} catalogue to 6000 sources (23\%). Finally the two samples are cross-matched with a large maximum distance of 100\arcsec. Sources that are farther apart than $10\times$ the median absolute deviation (MAD) of the offsets are removed in an iterative process. The final number of sources after the cross-match is 2770 (final MAD: 1.1\arcsec).

The final mean separation between selected sources in our catalogue and FIRST catalogue was found to be $-0.28\arcsec$ in RA and $-0.03\arcsec$ in Dec with relative standard deviations RA=$2.50\arcsec$ and Dec$=2.18\arcsec$ (see Fig.~\ref{fig:astrometry}). Given the small global offset between our catalogue and FIRST we did not correct for the shift.

\begin{figure}[ht!]
\centering
 \includegraphics[width=.5\textwidth]{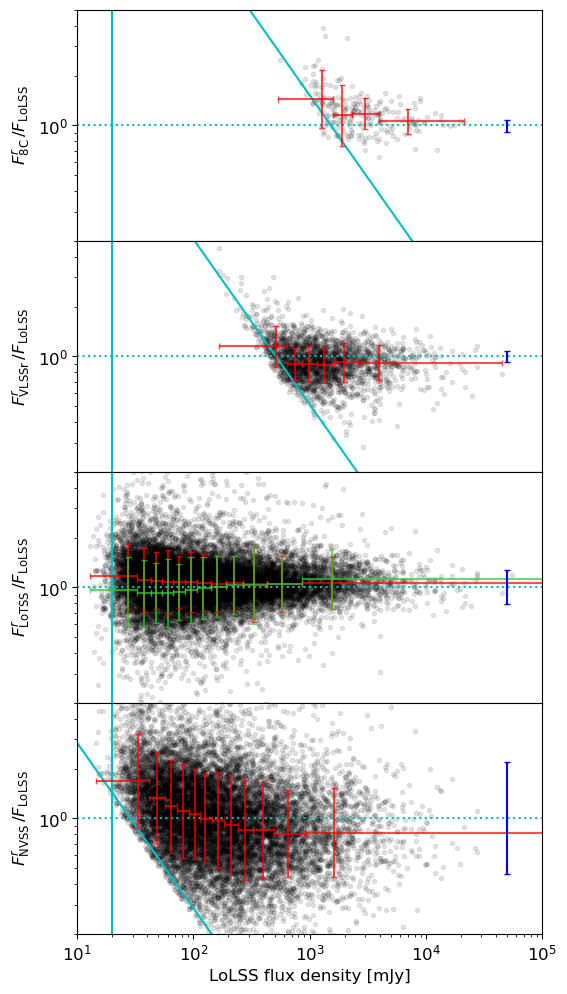}
 \caption{Ratio of the expected flux density extrapolated from other surveys over the flux density as measured in \lol{} as a function of flux density. From top to bottom, the surveys shown are 8C, VLSSr, LoTSS and NVSS. The extrapolated flux density is calculated assuming a spectral index $\alpha=-0.78$ (each source is a black circle). A ratio of 1 (dotted blue line) means perfect extrapolation of the flux density value. Solid lines are detection limits imposed by the survey depth, the vertical line is due to the \lol{} limit (assumed $1\sigma = 4$~\mjybeam), the diagonal line is the sensitivity limit of the survey used for comparison. Red crosses are centered on the binned medians and show the standard deviations on the y direction and the bin size in the x direction. Green crosses are the same but assuming a flux-dependent spectral index as found by \citet{deGasperin2018}. The dark blue lines show the expected dispersion due to the spectral index distribution.}
 \label{fig:flux}
\end{figure}

\subsection{Flux density uncertainties}
\label{sec:flux}

To calibrate direction-independent effects as well as the bandpass response of the instrument (see Sect.~\ref{sec:calibration}) we used one of the following flux calibrators: 3C\,196 (50\% of the observations), 3C\,295 (40\% of the observations), and 3C\,380 (10\% of the observations). The choice of the calibrator depends on the elevation of the source at the moment of the observation. The flux density of the calibration models was set according to the low-frequency models of \cite{Scaife2012} and it has a nominal error ranging between two and four percent depending on the source used.

The LOFAR LBA system is rather simple and stable: two-beam observations, pointing at two flux calibrators simultaneously, showed that the flux density of one could be recovered using the bandpass calibration from the other at the five percent level. We can use this value as an estimation of the flux density accuracy. Within the presented survey area, there is also 3C\,295, whose flux density can be measured at the end of the calibration and imaging process to assess whether it is consistent with the value given by \cite{Scaife2012}. In the final survey image, the integrated flux density of 3C\,295 is 130~Jy, against an expected flux density of 133~Jy ($\sim $ 2\%\  error). This can be used to establish an idea of the flux density precision. Adding in quadrature the nominal error on the flux scale (4\%) with these two errors provides a global error budget of 7\%.

To validate this estimation we can compare \lol{} flux densities with those from other surveys. This is not trivial as no surveys of sufficient depth to measure the spectral index of a meaningful number of sources in the survey footprint are available at frequencies lower than 54 MHz. This procedure can be attempted using the 8C survey at 38 MHz, although only 230 sources from 8C are visible in \lol{} due to the partial overlap of the surveys' footprints. The alternative approach to validate the flux level relies on extrapolating the flux densities down to \lol{} frequency from higher frequency surveys.

In order to double check the flux density calibration of our catalogue, we used data from 8C (38 MHz), VLSSr (74 MHz), LoTSS-DR2\footnote{LoTSS Data Release 2 will be presented in the forthcoming publication from Shimwell et al. (in prep.).} (144 MHz), and NVSS (1400 MHz). For each of these surveys, as well as for \lol,{} we restricted the catalogue to isolated sources as described in Sect.~\ref{fig:astrometry}, using a minimum distance between sources of two times the survey resolution. Each catalogue was then cross-matched with the \lol{} catalogue, allowing for a maximum separation of 6\arcsec{} (15\arcsec{} in the case of VLSSr and 60\arcsec{} for 8C). Because of ionospheric smearing, for \lol{} and LoTSS, we used the integrated value of the flux density. As a first test, we cross-checked the flux density value of 3C\,295 with that expected from \citet{Scaife2012}. All surveys covering 3C\,295, except LoTSS, appear to be consistent within a few percent with the expected flux, as shown in Table~\ref{tab:3c295}. Dynamic range limitations seem to have affected the LoTSS image quality in that region.

As a next test, we rescaled the flux density of each survey to the expected value at 54 MHz assuming a flux-independent spectral index of $\alpha=-0.78$ \citep{deGasperin2018}. The standard deviation of the spectral index distribution is rather large ($\sigma = 0.24$) and implies a large scatter of the rescaled values, mostly when extrapolating from NVSS data. The results are presented in Fig.~\ref{fig:flux} (red crosses). Caution must be used when interpreting these plots as the limited sensitivity of the other surveys can bias the result, predicting higher than real flux densities for faint sources. That is the case for VLSSr, as well as for NVSS, where the surveys are not deep enough to sample the faint and steep-spectrum sources present in \lol{}. Given its lower frequencies, 8C will instead miss faint, flat spectrum sources. The diagonal blue lines in Fig.~\ref{fig:flux} predict these cutoff levels, which are more relevant the shallower the reference survey and the larger its frequency distance from 54 MHz (slope of the line). This is not a problem for LoTSS, where the depth is sufficient such that the great majority of the sources (up to a spectral index of $-4$) can be sampled.

Another bias comes from the assumption of the spectral index being flux-independent. In \cite{deGasperin2018}, the authors showed that the average $150-1400$ MHz spectral index has a non-negligible dependence on the flux density of the selected source, with fainter sources having flatter spectra. Using the flux-dependent spectral index values tabulated by \cite{deGasperin2018} for flux densities derived at 150 MHz, we can rescale LoTSS flux densities more precisely to the expected values at 54 MHz. The results are presented in Fig.~\ref{fig:flux} with green crosses. For LoTSS, where no bias for the survey depth is present, the average flux ratio between the flux densities rescaled to 54 MHz ($F^r$) and the \lol{} flux densities is $F^r_{\rm LoTSS}/F_{\rm LoLSS} = 0.99$ (with a flux-independent spectral index it is $F^r_{\rm LoTSS}/F_{\rm LoLSS} = 1.08$). A single spectral index scaling gives good predictions both for 8C (with $F^r_{\rm 8C}/F_{\rm LoLSS} = 1.05$ in the brightest bin) and for VLSSr matched sources (with $F^r_{\rm VLSSr}/F_{\rm LoLSS} = 0.91$ in the brightest bin).

\begin{figure}
\centering
 \includegraphics[width=.5\textwidth]{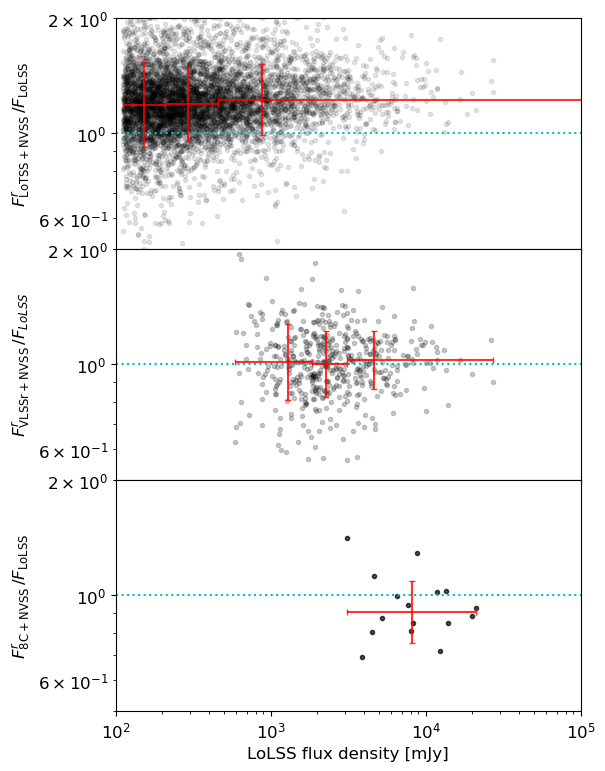}
 \caption{Ratio of the expected flux density derived from the combination of two other surveys over the flux density as measured in \lol{} as a function of flux density in \lol{}. In this case, the expected flux density is extrapolated using a spectral index derived from the combination of the following surveys: LoTSS-NVSS (top), VLSSr-NVSS (middle), and 8C-NVSS (bottom).}
 \label{fig:flux_spidx}
\end{figure}

\begin{figure}
\centering
 \includegraphics[width=.5\textwidth]{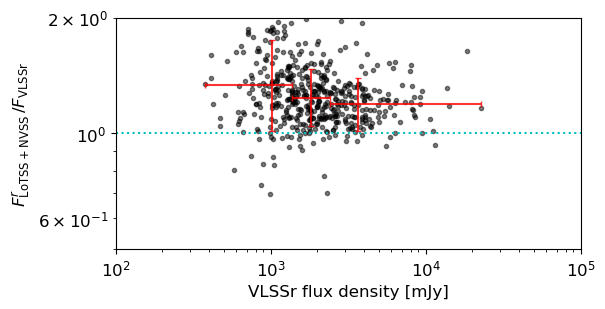}
 \caption{Same as in Fig.~\ref{fig:flux_spidx} but here NVSS and LoTSS are used to predict the flux density of VLSSr, obtaining a similar level of overprediction.}
 \label{fig:flux_spidx_vlss}
\end{figure}

A way to circumvent the assumption of using a single spectral index for different sources is to extract the spectral index value directly from two surveys and interpolate or extrapolate the flux density to 54 MHz. In Fig.~\ref{fig:flux_spidx}, we show how this approach systematically overestimates, by about 20\%, the expected \lol{} flux density when using NVSS and LoTSS to estimate the spectral index of the sources. On the other hand, using a survey closer in frequency, such as VLSSr, drastically reduces the effect. This is visible from the second panel of Fig.~\ref{fig:flux_spidx}, where the average ratio between the extrapolated flux density and that measured in \lol{} is 1.01. Also, the interpolation between 8C and NVSS predicts the \lol{} flux density with an average accuracy of 6\%, but based on only 45 sources.

As a final cross-check we also tried to predict the flux densities of VLSSr sources using LoTSS and NVSS data (Fig.~\ref{fig:flux_spidx_vlss}). We found that the predicted flux is overestimated by about 25\%. This is another way to confirm that the \lol{} and VLSSr flux scales are in agreement, whilst it shows a disagreement between the \lol{} and LoTSS flux scales. However, this approach has two limitations: each source needs to be detected in three surveys, reducing the total number of sources, and it relies on the assumption of a pure power law extrapolation. The latter is not a good assumption at 54 MHz, where a number of sources experience a curvature of the spectrum \citep[e.g.][]{deGasperin2019}. However, the fraction of sources with a curved spectrum is expected to be on the order of $\sim20-30$\% \citep{Callingham2017}, which should only affect that fraction of sources above the ratio $=1$ line in the top panel of Fig.~\ref{fig:flux_spidx} and, thus, should not create the global offset that we found. One possibility is that \lol{} and VLSSr are both offset towards lower flux densities by the same amount (up to $\sim20\%$). This seems an improbable coincidence and would contradict the good results obtained on the flux scale tests on calibrators and the good agreement with the 8C+NVSS interpolation. Alternatively, the 6C survey, on which LoTSS flux densities are rescaled \citep{Hardcastle2020}, might be offset (towards higher flux densities) by $\sim 10\%$, but this also seems unlikely since 6C used observations of Cygnus A to calibrate and so should be by construction on the flux scale of \cite{Roger1973}. We note that the accuracy (i.e. a global offset) of 6C is estimated to be within $\pm 5\%$ \citep{Hales1988}, whilst the accuracy of LoTSS is estimated to be $\sim 10\%$ \citep{Shimwell2019}. Taking into account these different tests, we cannot derive a more conclusive estimate of the flux density accuracy, but it is reasonable to suggest that assuming a conservative 10\% error on the \lol{} flux density scale could be beneficial.

\begin{table*}
\caption{Measured flux densities for 3C\,295 in various surveys and the expected value following \citet{Scaife2012}.}\label{tab:3c295}
\centering
\begin{threeparttable}
\begin{tabular}{lcccc}
\hline\hline
Survey & Frequency & Measured & Expected & Fractional error\\
name & (MHz) & flux density (Jy) & flux density (Jy) & (per cent)\\
\lol{} & 54 & 129.7 & 133.3 & $-2.7$\\
VLSSr & 74 & 128.9 & 132.0 & $-2.3$\\
LoTSS & 144 & 81.3 & 100.1 & $-18.8$\\
NVSS & 1400 & 22.5 & 22.7 & $-0.9$\\
\hline
\end{tabular}
\end{threeparttable}

\end{table*}

\section{Public data release}
\label{sec:publicddatarelease}

The data presented in this paper are available online in the journal repository and online\footnote{\url{https://www.lofar-surveys.org/}} in the form of a source catalogue and a mosaic image. The image and catalogue cover a region of 740 deg$^2$. Of this region, around 500 deg$^2$ is covered at full depth, whilst the rest is located at the mosaic edges and therefore covered, with a reduced sensitivity.

\subsection{Source catalogue}
\label{sec:catalogue}

The catalogue contains 25,247 sources. Although we used an adaptive rms box size, a few artefacts around bright sources might still be present, and no attempt has been made to remove them. The catalogue retains the type of source as derived by PyBDSF, where it distinguishes isolated compact sources (source\_code = 'S'), large complex sources (source\_code = 'C'), and sources that are within an island of emission that contains multiple sources (source\_code = 'M'). 

We note that the catalogue may contain some blended sources, although the chance of this is low given the sky density. No attempt has been made to correct the PyBDSF catalogue into physical radio sources \citep[cf.][]{Williams2019}. Furthermore, we note that the uncertainties on the source position and on the flux density are derived locally by the source finder from the images and do not include the other factors discussed in the previous Sections. The most conservative approach is to add 2.5\arcsec{} (see Sect.~\ref{sec:astrometry}) in quadrature to the position error and 10\% of the flux density in quadrature to the flux error (see Sect.~\ref{sec:flux}). An extract of the catalogue is presented in Table~\ref{tab:catalogue-example}.

\begin{table*}
\centering
\caption{Example of entries in the source catalogue. The entire catalogue contains 25,247 sources. The entries in the catalogue include: source name, J2000 right ascension (RA), J2000 declination (Dec), peak brightness ($\rm{S_{peak}}$), integrated flux density ($\rm{S_{int}}$), and the uncertainties on all of these values. The catalogue also contains the local noise at the position of the source (rms noise), and the type of source (where `S' indicates an isolated source which is fit with a single Gaussian; `C' represents sources that are fit by a single Gaussian but are within an island of emission that also contains other sources; and `M' is used for sources which are extended and fitted with multiple Gaussians). Not listed here, but present in the catalogue, there is also the estimation of the source size, both with and without the effect of beam convolution. The uncertainties on source positions and the flux densities are derived locally by the source finder and are likely underestimated (see text).} \label{tab:catalogue-example}
\begin{tabular}{ccccccccccc}
\hline
Source name & RA & $\sigma_{RA}$ & DEC & $\sigma_{DEC}$ & $S_{peak}$ & $\sigma_{Speak}$  & $S_{int}$ & $\sigma_{Sint}$ & rms noise & Type\\
& ($^\circ$)  & ($\arcsec$) & ($^\circ$)  & ($\arcsec$) & (mJy/beam) & (mJy/beam) & (mJy)  &  (mJy) & (mJy/beam) & \\ 
\hline
LOLpJ110902.0+571931 & 167.258 & 2.4 & 57.325 & 2.4 & 30.8 & 6.7 & 34.5 & 4.2 & 4 & S \\
LOLpJ110903.2+515046 & 167.263 & 0.4 & 51.846 & 0.6 & 691.4 & 27.2 & 396.8 & 6.8 & 7 & M \\
LOLpJ110903.3+525540 & 167.264 & 5.0 & 52.928 & 3.9 & 115.1 & 21.9 & 59.2 & 7.8 & 7 & S \\
LOLpJ110903.4+514027 & 167.264 & 1.6 & 51.674 & 1.3 & 166.6 & 15.0 & 117.8 & 6.8 & 6 & S \\
LOLpJ110904.3+592725 & 167.268 & 5.3 & 59.457 & 5.8 & 57.2 & 15.5 & 35.6 & 6.3 & 6 & S \\
LOLpJ110905.1+551619 & 167.271 & 0.2 & 55.272 & 0.2 & 2633.3 & 42.0 & 2263.0 & 21.9 & 21 & S \\
LOLpJ110905.2+460200 & 167.272 & 5.8 & 46.033 & 1.8 & 161.4 & 23.6 & 60.9 & 6.8 & 7 & M \\
LOLpJ110906.1+580944 & 167.275 & 2.8 & 58.162 & 2.5 & 51.0 & 9.0 & 40.4 & 4.4 & 4 & S \\
LOLpJ110906.2+474809 & 167.276 & 4.7 & 47.802 & 3.7 & 31.8 & 9.2 & 26.2 & 4.6 & 4 & S \\
LOLpJ110906.4+513040 & 167.277 & 0.6 & 51.511 & 0.7 & 404.6 & 23.4 & 230.2 & 6.2 & 6 & M \\
LOLpJ110909.1+512407 & 167.288 & 3.2 & 51.402 & 2.3 & 112.8 & 16.2 & 66.2 & 6.4 & 6 & S \\
LOLpJ110909.2+530019 & 167.288 & 5.9 & 53.005 & 6.0 & 70.8 & 19.6 & 38.2 & 7.2 & 7 & S \\
LOLpJ110910.2+494005 & 167.293 & 3.8 & 49.668 & 3.3 & 37.6 & 8.7 & 28.8 & 4.2 & 4 & S \\
LOLpJ110912.2+594151 & 167.301 & 5.4 & 59.698 & 4.1 & 41.3 & 12.7 & 33.7 & 6.3 & 6 & S \\
LOLpJ110912.6+532850 & 167.303 & 1.4 & 53.481 & 1.4 & 163.5 & 14.8 & 122.0 & 7.0 & 7 & S \\
LOLpJ110912.7+574619 & 167.303 & 4.2 & 57.772 & 3.5 & 36.9 & 8.9 & 27.0 & 4.1 & 4 & S \\
LOLpJ110913.6+580031 & 167.307 & 1.2 & 58.009 & 1.0 & 118.3 & 8.9 & 94.2 & 4.4 & 4 & S \\
LOLpJ110913.9+570756 & 167.308 & 0.3 & 57.132 & 0.3 & 457.8 & 8.8 & 351.3 & 4.2 & 4 & S \\
LOLpJ110914.0+542731 & 167.308 & 5.2 & 54.459 & 4.7 & 47.0 & 15.1 & 35.0 & 7.1 & 7 & S \\
LOLpJ110914.1+570936 & 167.309 & 4.1 & 57.160 & 3.5 & 31.7 & 8.5 & 25.8 & 4.3 & 4 & S \\
\end{tabular}
\end{table*}

We estimate the completeness of the catalogues following the procedure outlined by \cite{Heald2015}. For this process we used the residual mosaic image created after subtracting the sources detected by PyBDSF. This image carries the information of the distribution of the rms noise of the real mosaic and can therefore be used to inject fake sources and assess to what level they can be retrieved. We inject a population of 6000 point sources, randomly distributed, with flux densities ranging between 1 mJy and 10 Jy, and following a number count power-law distribution of $\frac{dN}{dS}\propto S^{-1.6}$. To simulate ionospheric smearing, the peak flux density of each source is reduced by 20 percent, whilst its size is increased to preserve the integrated flux density. We then attempt the detection of these sources using PyBDSF with the same parameters used for the catalogue. The process is then repeated 50 times to decrease sample noise.

\begin{figure}
\centering
 \includegraphics[width=.5\textwidth]{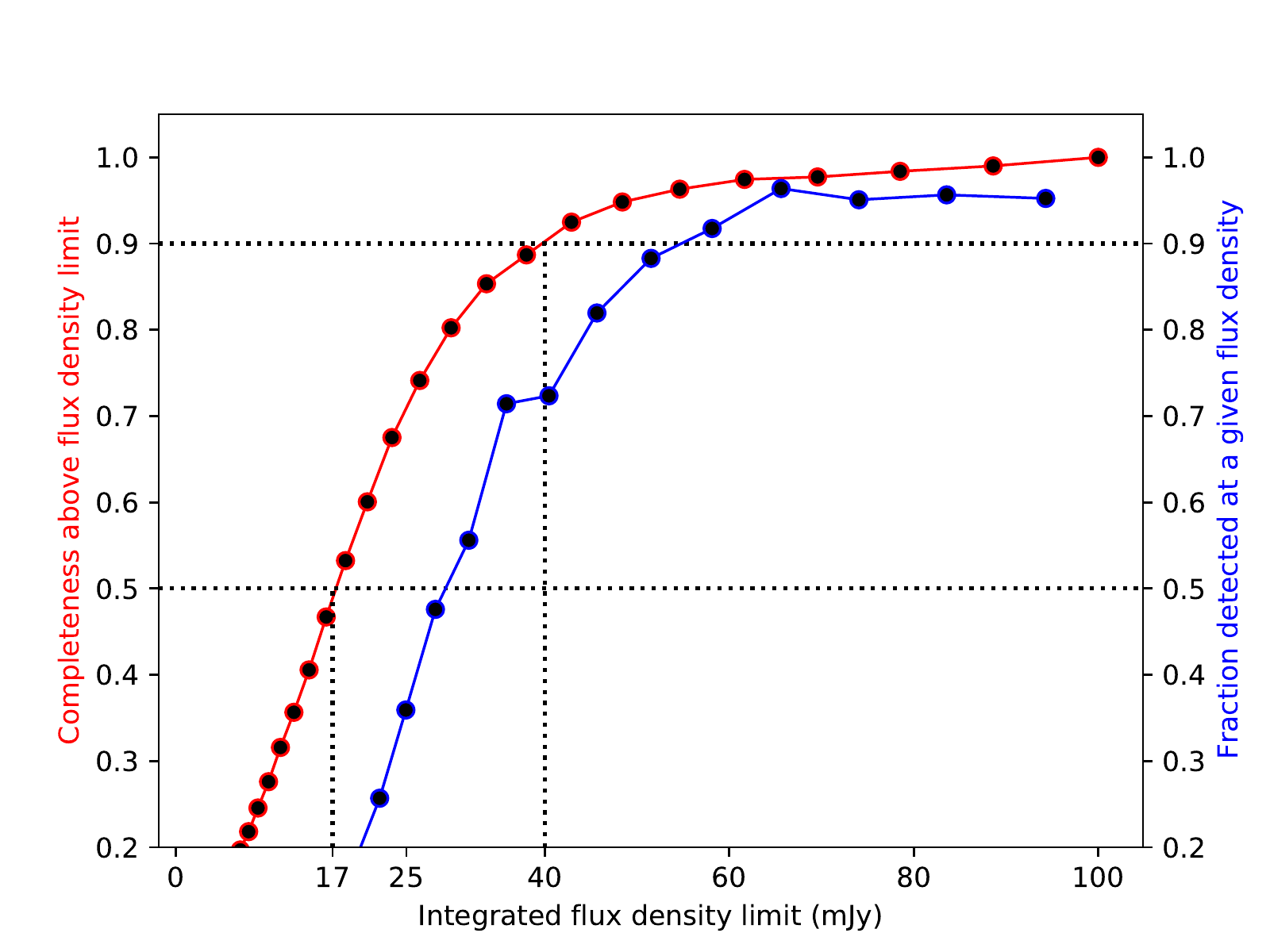}
 \caption{Estimated cumulative completeness of the preliminary data release catalogue (red) and the fraction of simulated sources that are detected as a function of flux density (blue), both assuming $\frac{dN}{dS}\propto S^{-1.6}$.}
 \label{fig:completeness}
\end{figure}

We consider a source as detected if it is found to be within 25\arcsec{} of its input position and with a recovered flux density that is within ten times the error on the recovered flux density from the simulated value. We found that we have a 50 percent probability of detecting sources at 25~mJy and 90 per cent probability of detecting sources at 50 mJy. In Fig.~\ref{fig:completeness}, we show the completeness over the entire mosaiced region (740 deg$^2$), that is, the fraction of recovered sources above a certain flux density. Our simulations indicate that the catalogue is 50 percent complete over 17 mJy and 90 per cent complete over 40 mJy, although we note that these values for cumulative completeness depend on the assumed slope of the input source counts.

The mosaic image has about 10$^8$ valid pixels, that is the region where at least one primary beam attenuation was higher than 30\%. In the case of pure white noise, with a $5\sigma$ detection limit we expect around 100 false positives. However, the background noise of the mosaic image is largely dominated by systematic effects. To assess the number of false positives, we started from the mosaic image used to build the catalogue and we invert its pixel values. Negative pixels due to noise and artefacts are now positive, whilst its sources are negative. Running the source finder with the same parameters used in the original run, we evaluated how many artefacts are erroneously considered legitimate sources. During this process, we used the same rms mask produced for the original detection because that evaluation is influenced by the positive pixels. We detected 1055 sources, highly concentrated along the mosaic edges and around bright sources. From this, we conclude that the number of false positive in our catalogue is around 4\%.

\begin{figure*}[htb!]
 \centering
 \includegraphics[angle=90,height=24cm]{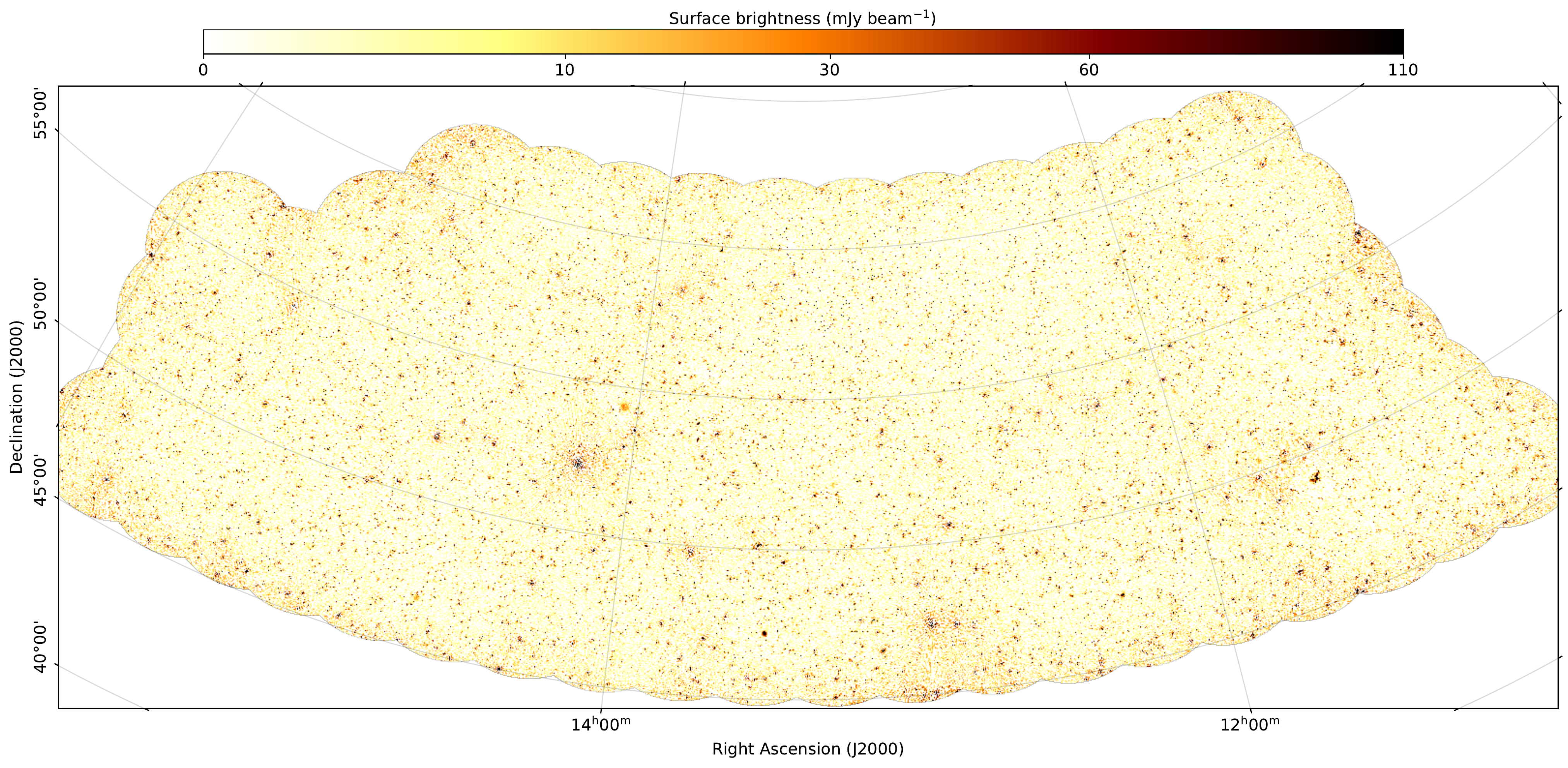}
 \caption{Mosaic image of the preliminary release of \lol{}, covering the HETDEX spring field region. Beam size: \beam{47}{47}.}
 \label{fig:mosaic}
\end{figure*}

\begin{figure}
\centering
 \includegraphics[width=0.5\textwidth]{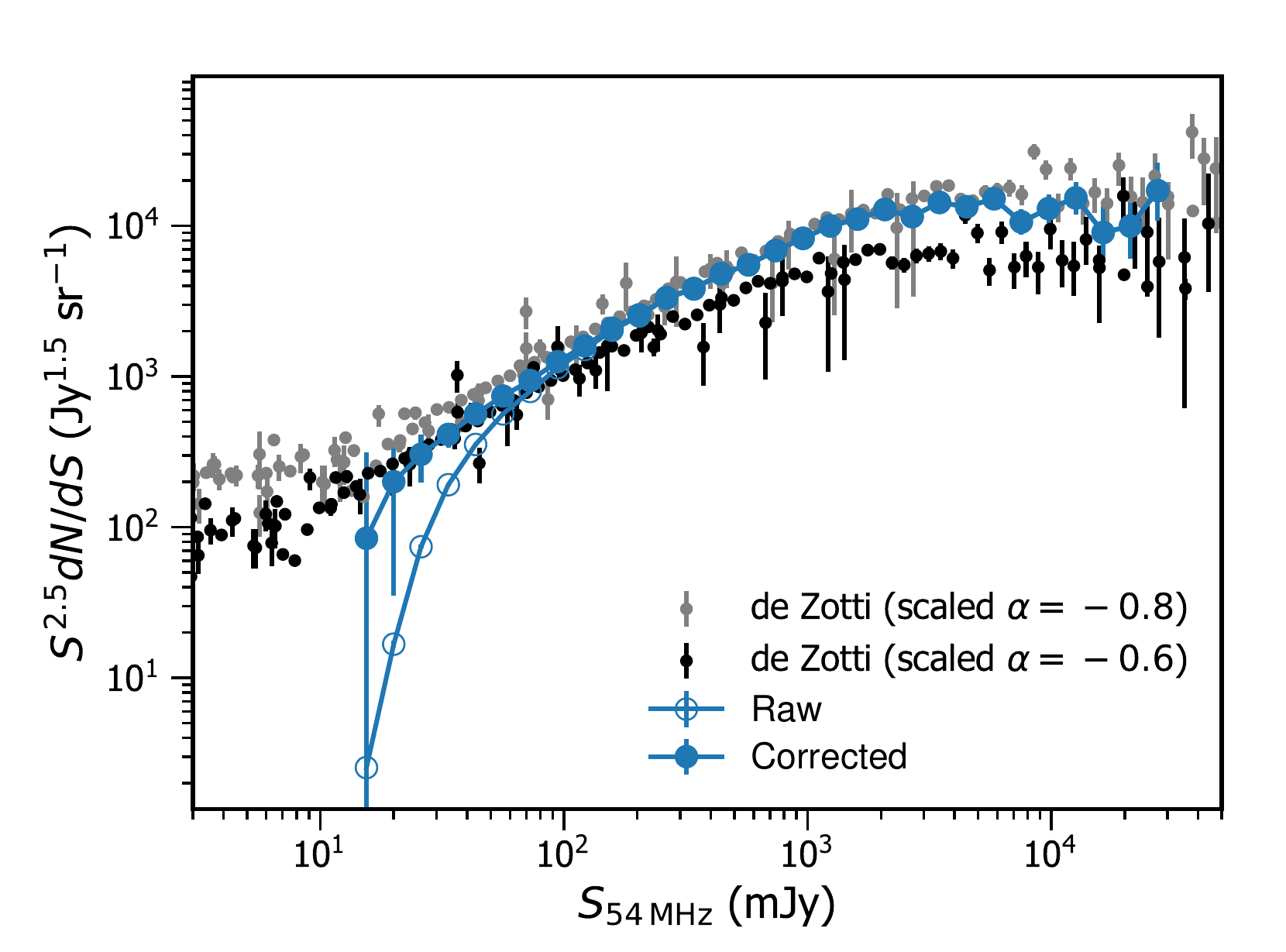}
 \caption{Euclidean-normalised differential source counts for \lol{} between 10\,mJy and 30\,Jy. The open circles show the raw, uncorrected source counts, whilst the filled circles show the counts corrected for completeness. For comparison, we show the 1.4\,GHz source counts from various surveys compiled \citet{deZotti2010}, and scaled to 54\,MHz,  assuming a spectral index of -0.8 (in gray) and -0.6 (in black).}
 \label{fig:srccnt}
\end{figure}

We calculated the Euclidean-normalised differential source counts for the \lol{} catalogue presented here. These are plotted in Fig~\ref{fig:srccnt}. Uncertainties on the final normalised source counts were propagated from the error on the completeness correction and the Poisson errors \citep{Gehrels1986} on the raw counts per flux density bin. To account for incompleteness, we used the measured peak intensities to calculate the fractional area of the survey in which each source could be detected, $A_i$. The count in each flux density bin is then determined as $N = \sum {1/A_{i}}$. To estimate an error on this correction, we  used the measured uncertainty on each peak intensity to determine an error in the visibility area of each source. The resolution bias, which takes into account the size distribution of sources and non-detection of large sources, is negligible at this resolution. For comparison we considered the  1.4\,GHz source counts compilation of \citet{deZotti2010}, scaled down to 54\,MHz assuming two different spectral indices. The \lol{} counts show good agreement with these previously determined counts, with a transition at around 100\,mJy of the average spectral index from $-0.8$ to $-0.6$ at lower flux densities. These values are consistent with the flux-dependent spectral index discussed in Sect.~\ref{sec:flux}.

\subsection{Mosaic image}
\label{sec:mosaic_image}

\begin{figure*}
\centering
 \includegraphics[width=.8\textwidth]{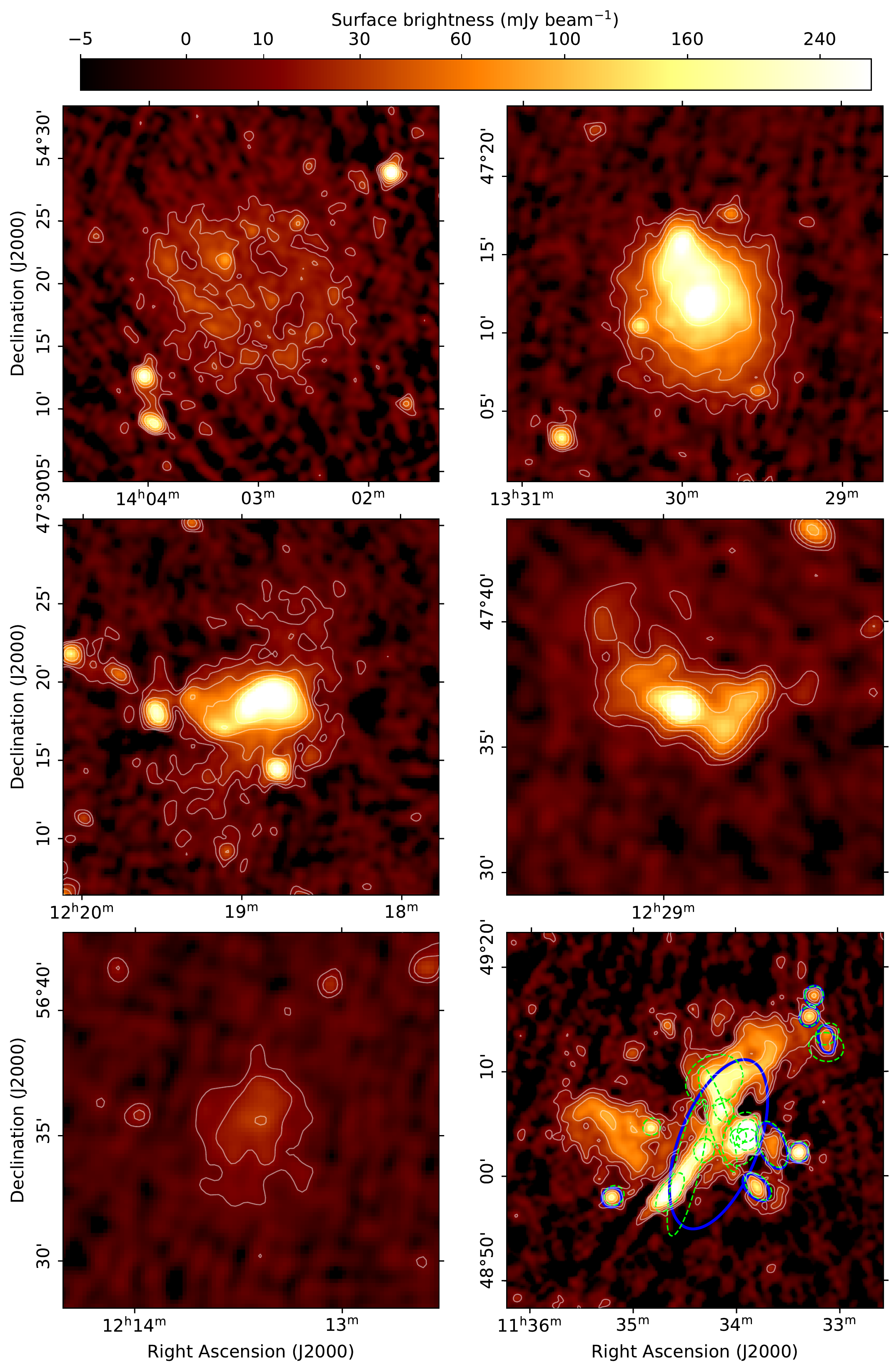}
 \caption{Some examples of extended sources in the data release published with this paper. From top-left to bottom-right, the nearby galaxies M101, M51, M106, the galaxy cluster Abell 1550, the emission surrounding the early type galaxy MCG+10-18-009, and the complex blend of emission coming from both AGN activity and diffuse sources in the ICM in Abell~1314. Contours start at three times the local rms noise. In the last panel, the green dashed lines show the gaussians used by the source finder to model the brightness distribution of the radio sources, the blue regions show the location and size of the sources, composed by one or more Gaussian components, as they are present in the catalogue.}
 \label{fig:extended}
\end{figure*}

The released image reveals the radio sky at 42--66 MHz with a depth of 4--5~\mjybeam and a resolution of 47\arcsec (see Fig.~\ref{fig:mosaic}). Even at a lower angular resolution and reduced sensitivity as compared to what will be achieved in the full survey, a number of nearby galaxies, radio galaxies, and galaxy clusters show the presence of resolved diffuse emission in the survey images. In Fig.~\ref{fig:extended}, we present examples of three nearby galaxies, two galaxy clusters, and faint diffuse emission surrounding an early type galaxy, probably hinting at past nuclear activity. We note that PyBDSF might not correctly associate all emission of the few very extended sources in the catalogue, as shown in the last panel of Fig.~\ref{fig:extended}.

\begin{figure*}
\centering
 \includegraphics[width=\textwidth]{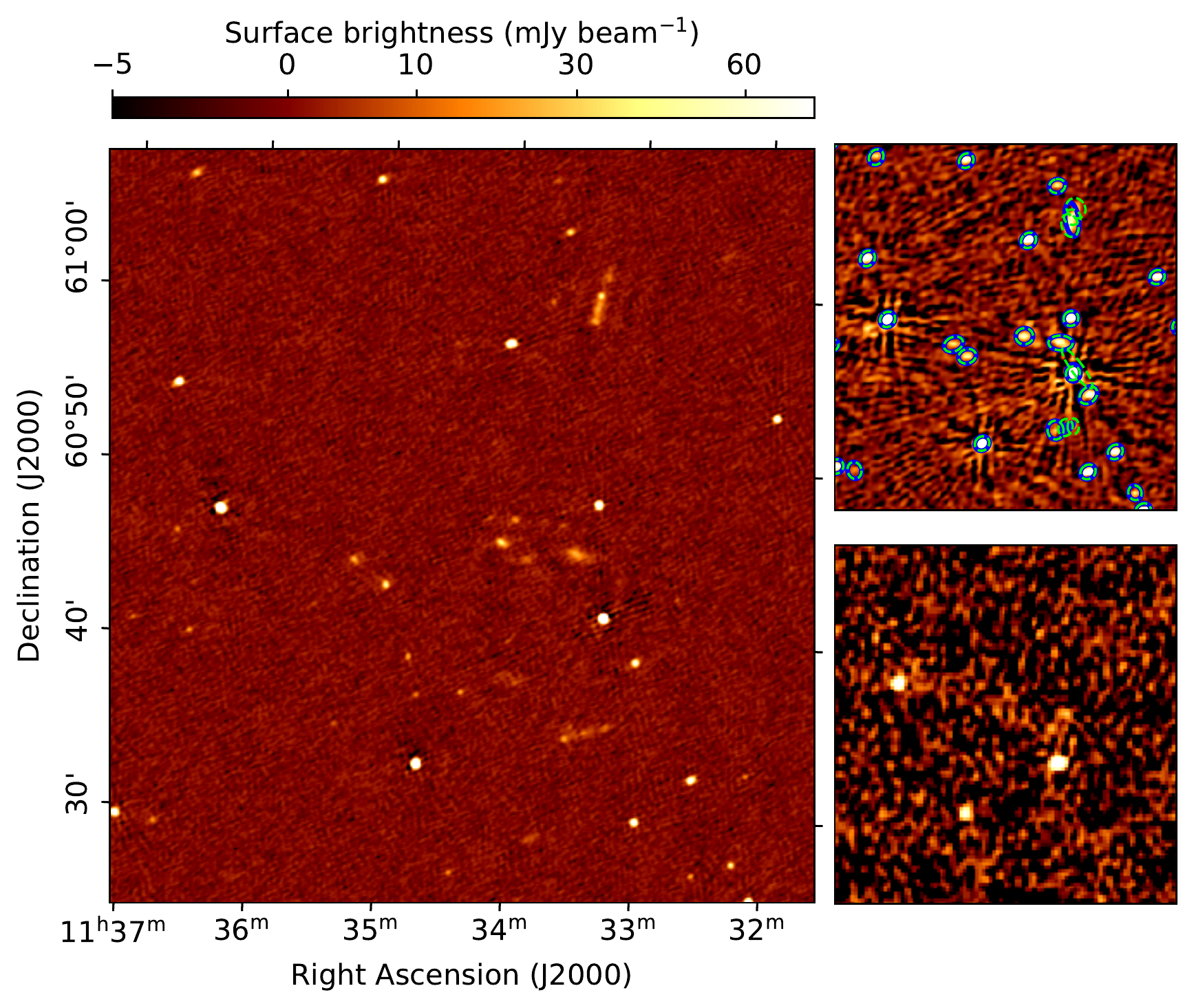}
 \caption{Region of the LOFAR LBA sky survey re-imaged using direction-dependent calibration (rms noise: 1.3~\mjybeam{} - beam: \beam{15}{15}). Top right: Same field without direction-dependent calibration (as presented in this paper; rms noise: 3~\mjybeam{} - beam: \beam{46}{32}). Green and blue regions show the location of the identified sources as described in~\ref{fig:extended}. Bottom right:  Same field in the VLSSr survey (rms noise: 73~\mjybeam{} - beam: \beam{80}{80}).}
 \label{fig:dd}
\end{figure*}

The other large-area survey that explores similar frequencies at comparable resolution of the preliminary release of \lol{} is VLSSr. That survey reaches an average rms map sensitivity of 130~\mjybeam at a resolution of 80\arcsec{} and a frequency of 74~MHz. A comparison between VLSSr and \lol{} is presented in Fig.~\ref{fig:dd}. The number of sources detected by \lol{} is around a factor of ten higher. However, the fidelity of extended sources as well as the noise level in the vicinity of the brightest sources are still compromised by a missing direction-dependent correction. Therefore, we warn the reader of possible larger errors in the flux density of extended sources compared to what is estimated in the paper. The effects of direction-dependent correction will mitigate this problem. We tested the results of such a procedure, reprocessing a few pointing of \lol{} with the experimental direction-dependent correction strategy outlined in \citet{deGasperin2020a}. The resulting image is presented in the large panel of Fig.~\ref{fig:dd}, where the final resolution is 15\arcsec{} and the noise approaches the thermal noise at 1.3~\mjybeam. Compared to the direction-independent calibrated image, 30\% more sources are detected in the direction-dependent calibrated image, with a superior image fidelity. This illustrates the potential of the final release of \lol. A full analysis of all the fields reprocessed with direction-dependent calibration will be presented in a forthcoming paper.

Despite the lower angular resolution and lower sensitivity of the released images, a number of projects from those described in Sect.~\ref{sec:science_cases} can still be carried out. Examples include the estimation of CRe diffusion in face-on galaxies such as M51 (second panel of Fig.~\ref{fig:extended}, Heesen et al., in prep.) or the search for emission from nearby exoplanets. Large-scale sources in dense environments, such as the nearby galaxy cluster Abell 1314 (last panel, Fig.~\ref{fig:extended}), point towards the presence of a large amount of diffuse steep spectrum emission; whilst a careful cross-match with other radio surveys and high-energy catalogues show the potential of \lol{} in the characterisation of blazars (Kadler et al in perp.).

\section{Summary}
\label{sec:summary}

In this work, we present the preliminary release of 740 deg$^2$ (95 pointings) of the LOFAR LBA Sky Survey. The data were processed using the Pipeline for LOFAR LBA (PiLL) code up to the point where the correction of direction-independent errors is complete. The final sensitivity of the preliminary release is 4--5~\mjybeam{} at a resolution of 47\arcsec. The catalogue that accompanies the paper contains more than 25,000 radio sources (detected with a $5\sigma$ rms threshold). We used Monte-Carlo simulations to assess the completeness of the catalogue and we conclude that it is 50\% complete for sources above a flux density of 17~mJy and 90\% complete for sources above a flux density of 40~mJy. We evaluated our astrometric accuracy to be within 2.5\arcsec{} ($1\sigma$). We cross-checked the flux density of the sources in our catalogue with other surveys and found good agreement on interpolated and extrapolated values. We estimate the flux density scale uncertainty to be within 10\%. The data presented in this paper, as well as the final survey products, are available to the community for scientific exploitation. 

The final aim of \lol{}, which we have shown here to be an achievable goal, is to cover the entire northern sky in the frequency range 42--66 MHz, at a sensitivity of $\sim 1$~\mjybeam{} and a resolution of 15\arcsec{} at optimal declination. The full survey will require 3170 pointings; currently, all pointings above a declination of 40\deg{} are being observed and these observations should be completed by mid-2022. We plan to further increase the coverage to a declination of 20\deg{} and, finally, to a declination of 0\deg{}. The final release of the survey will include a full direction-dependent error correction as demonstrated in Fig.~\ref{fig:dd} and \cite{deGasperin2020a}. 

The final release of the survey will facilitate advances across a range of astronomical research areas, as described in this work (see Sect.~\ref{sec:science_cases}). Together with the higher frequency counterpart at 144 MHz (LoTSS), \lol{} will allow for the study of more than 1 million low-frequency radio spectra, providing unique insights on physical models for galaxies, active nuclei, galaxy clusters, and other  fields of research. This experiment represents a unique attempt to explore the ultra-low frequency sky at a high angular resolution and depth. Thanks to its optimal combination of  resolution and sensitivity, the LOFAR LBA Sky Survey will remain unique well into the Square Kilometre Array (SKA) era.

\begin{acknowledgements}

LOFAR is the LOw Frequency ARray designed and constructed by ASTRON. It has observing, data processing, and data storage facilities in several countries, which are owned by various parties (each with their own funding sources), and are collectively operated by the ILT foundation under a joint scientific policy. The ILT resources have benefited from the following recent major funding sources: CNRS-INSU, Observatoire de Paris and Universit\'e d’Orleans, France; BMBF, MIWF-NRW, MPG, Germany; Science Foundation Ireland
(SFI), Department of Business, Enterprise and Innovation
(DBEI), Ireland; NWO, The Netherlands; The Science and
Technology Facilities Council, UK; Ministry of Science and Higher Education, Poland; Istituto Nazionale di Astrofisica (INAF). This research has made use of the University of Hertfordshire high-performance computing facility (\url{https://uhhpc.herts.ac.uk/}) and the LOFAR-UK compute facility, located at the University of Hertfordshire and supported by STFC [ST/P000096/1].

FdG and MB acknowledge support from the Deutsche Forschungsgemeinschaft under Germany's Excellence Strategy - EXC 2121 “Quantum Universe” - 390833306. GB and RC acknowledge partial support from mainstream PRIN INAF "Galaxy clusters science with LOFAR". MJH acknowledges support from the UK Science and Technology Facilities Council (ST/R000905/1). RJvW acknowledges support from the ERC Starting Grant ClusterWeb 804208. AB acknowledges support from the VIDI research programme with project number 639.042.729, which is financed by the Netherlands Organisation for Scientific Research (NWO). KLE acknowledges financial support from the NWO through TOP grant 614.001.351. MH acknowledges funding from the European Research Council (ERC) under the European Union's Horizon 2020 research and innovation programme (grant agreement No 772663). PNB is grateful for support from the UK STFC via grant ST/R000972/1. IP acknowledges support from INAF under the SKA/CTA PRIN “FORECaST” and the PRIN MAIN STREAM “SAuROS” projects. AB acknowledges support from ERC Stg DRANOEL n. 714245 and MIUR FARE grant “SMS”. IP and MV acknowledge support from the Italian Ministry of Foreign Affairs and International Cooperation (MAECI Grant Number ZA18GR02) and the South African Department of Science and Technology's National Research Foundation (DST-NRF Grant Number 113121) as part of the ISARP RADIOSKY2020 Joint Research Scheme.

\end{acknowledgements}

\bibliographystyle{aa}
\bibliography{../bbtex/library}

\end{document}